\documentclass{article}
\usepackage{amsmath}[1996/11/01]
\usepackage{amssymb,amsthm,amsxtra}
\usepackage{epic,eepic}
\usepackage{epsfig}

\newtheorem{thm}{Theorem}
\newtheorem{cor}[thm]{Corollary}
\newtheorem{lem}[thm]{Lemma}
\newtheorem{prop}[thm]{Proposition}
\theoremstyle{remark}
\newtheorem*{rem}{Remark}

\theoremstyle{definition}

\numberwithin{equation}{section}
\numberwithin{thm}{section}

\DeclareMathOperator{\Exp}{\mathbb E}

\DeclareMathOperator{\sgn}{sgn} \DeclareMathOperator{\Prob}{\mathbb P}

\DeclareMathOperator{\Pf}{Pf}

\newcommand{\C}{\mathbb C}
\newcommand{\R}{\mathbb R}
\newcommand{\Z}{\mathbb Z}
\newcommand{\N}{\mathbb N}

\DeclareMathOperator{\FLEN}{F_N} \DeclareMathOperator{\ELEN}{E_N}
\DeclareMathOperator{\aLEN}{a_N} \DeclareMathOperator{\bLEN}{b_N}

\newcommand{\symmO}{O}

\newcommand{\mone}{m^{(1)}}
\newcommand{\mtwo}{m^{(2)}}
\newcommand{\vtwo}{v^{(2)}}

\newcommand{\vR}{v_R}
\newcommand{\VN}{V_0}
\newcommand{\mthree}{m^{(3)}}
\newcommand{\vthree}{v^{(3)}}

\newcommand\psymmU{%
\begin{picture}(1,1)(0,0)%
\allinethickness{0.5pt}%
\path(0,0)(0,1)(1,1)(1,0)(0,0)%
\end{picture}}
\newcommand\psymmUU{%
\begin{picture}(1,1)(0,0)%
\allinethickness{0.5pt}%
\path(0,0)(0,1)(1,1)(1,0)(0,0)%
\put(0.5,0.5){\makebox(0,0){$\cdot$}}%
\end{picture}}
\newcommand\psymmO{%
\begin{picture}(1,1)(0,0)%
\allinethickness{0.5pt}%
\path(0,0)(0,1)(1,1)(1,0)(0,0)%
\path(0,0)(1,1)%
\end{picture}}
\newcommand\psymmS{%
\begin{picture}(1,1)(0,0)%
\allinethickness{0.5pt}%
\path(0,0)(0,1)(1,1)(1,0)(0,0)%
\path(1,0)(0,1)%
\end{picture}}
\newcommand\psymmu{%
\begin{picture}(1,1)(0,0)%
\allinethickness{0.5pt}%
\path(0,0)(0,1)(1,1)(1,0)(0,0)%
\path(0,0)(1,1)%
\path(0,1)(1,0)%
\end{picture}}

\newbox\tsymmUbox
\newbox\tsymmUUbox
\newbox\tsymmObox
\newbox\tsymmSbox
\newbox\tsymmubox
\setbox\tsymmUbox =\hbox{\kern0.75pt\setlength{\unitlength}{6pt}\psymmU \kern0.75pt}
\setbox\tsymmUUbox=\hbox{\kern0.75pt\setlength{\unitlength}{6pt}\psymmUU\kern0.75pt}
\setbox\tsymmObox =\hbox{\kern0.75pt\setlength{\unitlength}{6pt}\psymmO \kern0.75pt}
\setbox\tsymmSbox =\hbox{\kern0.75pt\setlength{\unitlength}{6pt}\psymmS \kern0.75pt}
\setbox\tsymmubox =\hbox{\kern0.75pt\setlength{\unitlength}{6pt}\psymmu \kern0.75pt}

\newbox\symmUbox
\newbox\symmUUbox
\newbox\symmObox
\newbox\symmSbox
\newbox\symmubox
\setbox\symmUbox =\hbox{\kern0.75pt\setlength{\unitlength}{4.5pt}\psymmU \kern0.75pt}
\setbox\symmUUbox=\hbox{\kern0.75pt\setlength{\unitlength}{4.5pt}\psymmUU\kern0.75pt}
\setbox\symmObox =\hbox{\kern0.75pt\setlength{\unitlength}{4.5pt}\psymmO \kern0.75pt}
\setbox\symmSbox =\hbox{\kern0.75pt\setlength{\unitlength}{4.5pt}\psymmS \kern0.75pt}
\setbox\symmubox =\hbox{\kern0.75pt\setlength{\unitlength}{4.5pt}\psymmu \kern0.75pt}

\def\symmO{{\copy\symmObox}}

\begin{document}

\title{{\bf Painlev\'e expressions for LOE, LSE \\ 
and interpolating ensembles}}
\author{{\bf Jinho Baik}\footnote{
Deparment of Mathematics,
Princeton University, Princeton, New Jersey, 08544,
jbaik@math.princeton.edu}
\footnote{Institute for Advanced Study,
Princeton, New Jersey 08540}}


\date{May 6, 2002}
\maketitle

\begin{abstract}

We consider an ensemble which interpolates the Laguerre orthogonal ensemble 
and the Laguerre symplectic ensemble.
This interpolating ensemble was introduced earlier by the author and Rains
in connection with a last passage 
percolation model with a symmetry condition.
In this paper, we obtain a Painelev\'e V expression 
for the distribution of the rightmost particle of the interpolating ensemble. 
Special cases of this result yield the Painlev\'e V expressions 
for the largest eigenvalues 
of  Laguerre orthogonal ensemble
and Laguerre symplectic ensemble of finite size.

\end{abstract}


\section{Introduction}

Given a sequence $\xi=(\xi_1, \cdots, \xi_N)$, the Vandermonde product 
of $\xi$ is denoted by
\begin{equation}\label{e-Delta}
  \Delta_N(\xi) = \prod_{1\le i<j\le N}(\xi_i-\xi_j).
\end{equation}
For any real constant $A>-1/2$, 
we consider the probability density function defined by
\begin{equation}\label{e-density}
  p(\xi_1,\cdots,\xi_N ; A)=Z_{N,A}^{-1}
\Delta_N(\xi) \prod_{j=1}^N e^{-\frac12 \xi_j} e^{A(-1)^j\xi_j}
\end{equation}
on the ordered set $\{ 0\le\xi_N\le \cdots\le \xi_1 \}$, where
\begin{equation}
  Z_{N,A} = (A+\frac12)^{-N} \prod_{j=0}^{N-1}j!
\end{equation}
is the normalization constant. 
The main purpose of this paper is to express the
distribution of the rightmost `particle' $\xi_1$ 
in terms of a solution of the Painlev\'e V equation (see
Theorem \ref{thm-PV} below). 

There are two reasons that we are interested in the above
density. 
The first is that \eqref{e-density} interpolates the 
Laguerre orthogonal ensemble
and the Laguerre symplectic ensemble in the random matrix theory, 
as $A$ varies from $0$ to $+\infty$. 
Letting $A\to 0$ or $A\to\infty$ in the Painlev\'e expression, 
we verify in particular that the distribution function for 
the rightmost particle $\xi_1$ satisfies the Painlev\'e V equation 
for the Laguerre orthogonal and symplectic ensembles, respectively.
These are new results in the literature.
The second is that the above density function 
represents the probability distribution of a certain
last passage percolation model with a symmetry condition.
Indeed, the density \eqref{e-density} was introduced
in \cite{BR1} (see Remark 7.6.1) as a formula for 
the distribution of the last passage time in this percolation model.
We now discuss these two aspects of the above density function.


\subsection*{Interpolating ensemble}

We first discuss the connection to 
the Laguerre ensembles. Let $w(x)$ be a weight
function on $\R$ or on a subset of $\R$ which decays 
sufficiently fast as $x\to\pm\infty$. Consider
the density function
\begin{equation}\label{e-Ldensity}
  \frac1{Z} \Delta_N(\xi)^\beta
\prod_{j=1}^N w(\xi_j)
\end{equation}
on the set $\xi_N\le\cdots\le\xi_1$, 
where $\beta>0$ is fixed, and $Z$ is the
normalization constant. 
The ensemble with the special choice of weight function
\begin{equation}\label{e-Ldensity2}
  w(x)=x^\alpha e^{-\frac12 x} 1_{x\ge 0}
\end{equation}
is called the Laguerre orthogonal ensemble (LOE), Laguerre unitary
ensemble (LUE), and Laguerre symplectic ensemble (LSE)
for $\beta=1,2$ and $4$, respectively (see e.g. \cite{Mehta}). 
The Laguerre
ensembles are of basic interest in the multivariate analysis of statistics (see, e.g.
\cite{Johnstone}). Especially, the LOE with $\alpha=M-1-N$ represents the probability of
the principal components (i.e. the singular values)
of an $M\times N$ random matrix $X$ whose entries are independent
(real) Gaussian random variables of mean 0, variance 1.

Introduce the density function, for real $A$, given by
\begin{equation}\label{e-dens}
   p(\xi_1,\cdots,\xi_N ; A; w;\beta) :=Z_{N,A;w;\beta}^{-1}
\Delta_N(\xi)^\beta \prod_{j=1}^N w(\xi_j) e^{A(-1)^j\xi_j}
\end{equation}
on the set $\R^N_{ord}:= \{\xi_N\le \xi_{N-1}\le\cdots \le \xi_1\}$. 
This density function
generalizes and also interpolates the ensembles \eqref{e-Ldensity} for
different $\beta$'s in the following sense.

\begin{prop}\label{prop-inter}
When $A=0$,
\begin{equation}\label{e-inter0}
   p(\xi_1,\cdots,\xi_N ; 0; w;\beta) = Z_{N,0; w}^{-1}
\Delta_N(\xi)^\beta \prod_{j=1}^N w(\xi_j)
\end{equation}
with some constant $Z(N,0;w)$. Let $w(x)=e^{-V(x)}$ be a positive, 
$C^1$-function supported on a subset of $\R$ 
such that $V(x)\ge c_0|x|$ for some $c_0>0$ 
as $x\to \pm \infty$, and $\| V'\|_{L^\infty} \le C_0$. 
Assuming that $N$ is even, for a bounded uniformly continuous
function $f(\xi_1, \cdots, \xi_N)$, we have
\begin{equation}\label{e-interinfty}
\begin{split}
  & \lim_{A\to +\infty} \int_{\R^N_{ord}} f(\xi_1, \cdots, \xi_N)
  p(\xi_1,\cdots,\xi_N ; A; w;\beta)  d\xi_1\cdots d\xi_N \\
  &\qquad =
  Z_{N,\infty; w;\beta}^{-1}
\int_{\R^{N/2}_{ord}} g(\zeta_1,\cdots, \zeta_{N/2})
\Delta_{N/2}(\zeta)^{4\beta}
   \prod_{j=1}^{N/2} w(\zeta_j)^2 d\zeta_j,
\end{split}
\end{equation}
where $g(\zeta_1,\zeta_2, \cdots, \zeta_{N/2})
:= f(\zeta_1,\zeta_1, \zeta_2,\zeta_2,\cdots,
\zeta_{N/2}, \zeta_{N/2})$ and
\begin{equation}\label{e-Zinfty}
  Z_{N,\infty;w;\beta} := \int_{\R^{N/2}_{ord}}
\Delta_{N/2}(\zeta)^{4\beta}
\prod_{j=1}^{N/2} w(\zeta_j)^2 d\zeta_j.
\end{equation}
We also have for any $t\in\R$,
\begin{equation}\label{e-char}
  \begin{split}
  & \lim_{A\to +\infty} \int_{\R^N_{ord}\cap\{\xi_1\le t\}}
  p(\xi_1,\cdots,\xi_N ; A; w; \beta) d\xi_1\cdots d\xi_N \\
  &\qquad =
  Z_{N,\infty; w;\beta}^{-1} \int_{\R^{N/2}_{ord}\cap \{\zeta_1\le t\}}
  \Delta(\zeta)^{4\beta}
   \prod_{j=1}^{N/2} w(\zeta_j)^2 d\zeta_j.
\end{split}
\end{equation}
\end{prop}

The case $A=0$ is trivial from the
expression \eqref{e-dens}. 
An explanation for the change $\beta\to 4\beta$ in the case when $A\to+\infty$ 
is the following.
We first
note that the term involving $A$ in \eqref{e-dens} is
\begin{equation}
  e^{-A(\xi_1-\xi_2+\xi_3-\xi_4+\cdots)}.
\end{equation}
Since $\xi_j$'s are ordered, $\xi_1\ge \xi_2\ge \cdots\ge \xi_N$ ($N$ even), 
the term
$\xi_1-\xi_2+\xi_3-\xi_4+\cdots$ is always non-negative. 
Thus as $A\to +\infty$, we have
a non-trivial limit for \eqref{e-dens} 
only when $\xi_1=\xi_2$, $\xi_3=\xi_4$, $\cdots$ 
(the term $Z_{N,A;w;\beta}^{-1}$ grows polynomially in $A$ ; 
see Lemma \ref{lem-inter} below). 
If we set $\zeta_1:= \xi_1=\xi_2$, $\zeta_2:= \xi_3=\xi_4$, $\cdots$, 
simple algebra shows that
the Vandermonde term $\Delta_N(\xi)$ in \eqref{e-dens}
becomes $\Delta_{N/2}(\zeta)^4$ if one drops the terms 
$(\xi_{2k-1}-\xi_{2k})$, $1\le k\le N/2$,
which vanish.
(See also the Remark in Section \ref{sec-inter} below for changes 
$\beta\to k^2\beta$, $k\in\N$). 
The full
proof of this Proposition is given in Section \ref{sec-inter}.

In view of the Coulomb gas interpretation of random matrix theory 
(see e.g. \cite{Mehta}), 
the term involving $A$ in the
density \eqref{e-dens} represents the pairwise-attraction of particles, 
in addition to
the log repulsion given by $\Delta_N(\xi)$ and 
the ``external field'' $w$. 
As $A$ becomes large, the pairwise-attraction becomes stronger, 
and for each $j=1,\cdots, N/2$, the pair of particles $\xi_{2j-1}, \xi_{2j}$ 
gets closer, and eventually sticks together to form
one particle $\zeta_j$.

If we take $w(x)=e^{-\frac12x}$ on $x\ge 0$ and $\beta=1$ 
in the above Proposition,
\eqref{e-density} becomes the density for LOE \eqref{e-Ldensity2} 
with $\alpha=0$ when $A=0$, 
while the limit of \eqref{e-density} as $A\to +\infty$ is, 
with the modification $e^{-\frac12\xi_j}$ by $e^{-\xi_j}$, 
the density for LSE \eqref{e-Ldensity2} with $\alpha=0$.
Thus by taking $A=0$ and $A=\infty$, 
the Painlev\'e V expression (Theorem \ref{thm-PV} below) 
for the largest particle
of the interpolating ensemble \eqref{e-density} yields the
Painlev\'e V expressions for the largest eigenvalue 
of LOE and LSE (see Corollary
\ref{cor-LOE} below) with $\alpha=0$ for any \emph{finite} $N\in \N$.

In the random matrix context, there have been many works that 
express the various distributions of matrix ensembles 
in terms of differential equations 
(see e.g. \cite{vanMoerbeke} and the references therein).
Most of results are for the unitary ensemble ($\beta=2$), 
and there are relatively few results for $\beta=1$ and $\beta=4$ 
(\cite{TW2, ForRai99, Forrestersmall, ForresterW2, Johnstone, ForresterWPVI}).
For the Laguerre
ensembles, the probability distributions for $\beta=2$ are 
found to be expressible in
terms of Painlev\'e V equation (\cite{TWFred}) for any finite $N$, 
and in terms of the Painlev\'e II equation (\cite{TW1}) 
in the limiting case $N\to\infty$. 
On the other hand, for $\beta=1$, Johnstone (\cite{Johnstone}) 
recently analyzed the limiting case $N\to\infty$, and
obtained a Painlev\'e II expression. 
The Painlev\'e V expressions for LOE/LSE for
\emph{finite} $N$ in this paper are new. 
While this paper was being written, Forrester and
Witte obtained different formulas for the largest eigenvalues 
of LOE and LSE (\cite{ForresterWPVI}). 
Their formulas involve Painlev\'e III' systems instead of
Painlev\'e V.
The relationship between these two formulas is an intriguing 
question and remains unclear.

\subsection*{Last passage percolation 
with a symmetry condition and the totally asymmetric exclusion
process with symmetry condition}

As mentioned earlier, the density function \eqref{e-density} 
also arises in connection with a
last passage percolation model.
For $r>0$, let $e(r)$ denote the exponentially distributed random
variable with mean $r$ : the density function of $e(r)$ 
is $r^{-1}e^{-x/r}$ for $x>0$ and
$0$ for $x\le 0$. By $e(0)$ we understand the random variable 
identically equal to
$0$. Fix $\rho\ge 0$. 
To each site $(i,j)\in \Z_+^2$, we attach a random variable 
$u(i,j)$ taken as follows :
\begin{align}
  u(i,j) &\sim e(1), \qquad i<j, \\
  u(i,i) &\sim e(\rho), \\
\label{e-usymm}  u(j,i)&=u(j,i), \qquad i<j.
\end{align}
Except for the symmetry condition $u(i,j)=u(j,i)$, 
the random variables are independent. 
Note that the condition \eqref{e-usymm} implies the symmetry
of the configuration of random variables in $\Z_+^2$ about the line $y=x$. 
An up/right path $\pi$ is a collection of sites
$\{(i_k,j_k)\}_{k=1}^r$ satisfying $(i_{k+1},j_{k+1})-(i_k,j_k)=(1,0)$ or $(0,1)$. Let
$\Pi(N)$ be the set of up/right path $\pi$ from $(1,1)$ to $(N,N)$. Define the random
variable
\begin{equation}
  H^\symmO(N;\rho) =
\max\bigl\{ \sum_{(i,j)\in\pi} u(i,j) : \pi \in \Pi(N)\bigr\}.
\end{equation}
If one interprets $u(i,j)$ as the (random) passage time to pass through the site $(i,j)$,
$H^\symmO(N;\rho)$ is the last passage time to go from $(1,1)$ to $(N,N)$ along a directed
(up/right) path. 
The relation of $H^\symmO(N;\rho)$ to the 
above interpolating density function is the following.

\begin{prop}\label{prop-conti}(Remark 7.6.1 \cite{BR1})
We have
\begin{equation}\label{e-multiint}
\Prob( H^\symmO(N;\rho) \le x) =
\int_{0\le\xi_N\le\cdots\le\xi_1\le x}
p(\xi_1, \cdots, \xi_N ; \frac1\rho-\frac12 ) \prod_{j=1}^N d\xi_j.
\end{equation}
\end{prop}

This Proposition is stated without full proof in 
Remark 7.6.1 \cite{BR1}: a full proof if given 
in Section \ref{sec-conti} below.

Closely related is a one-dimensional interacting particle system, 
the totally
asymmetric simple exclusion process (TASEP) (\cite{Liggett, kurtj:shape}). 
TASEP is a continuous-time stochastic process
on the integer lattice $\Z$. At any time, each site is either occupied by a particle or
empty. 
If a particle is at a site $j$ and its right-hand-site $j+1$ is empty, 
the particle jumps to its right-hand-site 
after a random waiting time given 
by an exponential random variable of mean $1$.
Thus the particles move only to the right. 
The waiting time for the jumps is independent and identically distributed 
at each site and each (continuous) time. 
These rules describe the usual 
totally asymmetric simple exclusion process 
(see e.g. \cite{kurtj:shape}, \cite{SpohnP2}, \cite{Liggett}). 
In \cite{kurtj:shape}, Johansson showed that the TASEP with special
initial data (all negative sites are occupied and all non-negative 
sites are empty) can be mapped to the last passage percolation 
above \emph{without} the symmetry condition \eqref{e-usymm}. 
This mapping was further generalized in \cite{SpohnP2} for TASEP with random
Bernoulli initial data.

The above last passage percolation model 
with the symmetry condition \eqref{e-usymm} 
is also related to a TASEP, but now the process
is defined only on the
non-negative integer lattice, $\N_0=\N\cup\{0\}$. 
For the sites $j>0$, the jump rules remain the same as before, but 
we assume that there is a
creation process at the origin $j=0$ : when the site $j=0$ is empty, a particle is created
after a random exponential waiting time of mean $\rho$. 
For initial data, we assume that all the sites are empty.
Then one can show that 
(\cite{SpohnP2}) the number of particles $N(t)$ that have been created at the
origin up to time $t$ satisfies
\begin{equation}
  \Prob ( N(t) \le N)  = \Prob( H^\symmO(N;\rho) > t).
\end{equation}

In the last passage percolation model above, one might also be interested in the last
passage time from $(1,1)$ to $(M,N)$ for general $M\neq N$. In terms of TASEP, this is
equivalent to the number of particles that have jumped across the site $M-N$. But there is
yet no formula like Proposition \ref{prop-conti} for general $M\neq N$.

\subsection*{Results}

Now we state the main results. 
From Proposition \ref{prop-conti}, 
the following results imply the Painlev\'e V expressions 
for the distribution of the rightmost `particle' 
from the interpolating ensemble \eqref{e-density}.

\begin{thm}\label{thm-PV}
With the notation $A=\frac1{\rho}-\frac12$, and $w=\frac2{\rho}-1=2A$, we have for $x>0,
\rho>0, N\in \N$,
\begin{equation}\label{e-thmPV}
\Prob(H^\symmO(N;\rho)\le x) =
\frac12\biggl\{\bigl[\aLEN(x,\rho)-\bLEN(x,\rho)\bigr] \bigl(\ELEN(x)\bigr)^{-1} +
\bigl[\aLEN(x,\rho)+\bLEN(x,\rho)\bigr]
\ELEN(x) \biggr\}\FLEN(x)
\end{equation}
for some functions
\begin{align}
 \FLEN(x) &:= \exp\biggl\{\int_x^\infty \frac14\alpha(y;N)dy\biggr\},\\
\label{e-Edefine} \ELEN(x) &:= \exp\biggl\{\int_x^\infty \frac14\beta(y;N)dy\biggr\}
\end{align}
and $\aLEN(x,\rho), \bLEN(x,\rho)$.
The functions $\alpha(x), \beta(x), \aLEN(x,\rho), \bLEN(x,\rho)$ 
are analytic in $x>0, \rho>0$, and 
satisfy the following properties :
\begin{itemize}
\item[(i)] (Painlev\'e V) The functions $\alpha(x;N), \beta(x;N)$ as functions in $x$ satisfy
\begin{equation}
  \alpha'(x;N)=\frac12 (\beta(x;N))^2.
\end{equation}
The function $\theta(x)=\theta(x;N):= -\frac12 x\alpha(x)$ solves the Painlev\'e V equation
\begin{equation}\label{e-painV}
 (x\theta'')^2=(\theta-x\theta')(\theta-x\theta'+4(\theta')^2+4N\theta').
\end{equation}
\item[(ii)] (Asymptotics of $\alpha, \beta$ as $x\to\infty$)
Fix $0<\epsilon<1/4$. For each fixed $N\in\N$, as $x\to +\infty$,
\begin{equation}\label{e-betaasymp1}
  \beta(x)
= 2(-1)^NL^{(1)}_{N-1}(x)e^{-\frac12x} + O(e^{-(1-\epsilon)x}) =
\frac{-2x^{N-1}}{(N-1)!}e^{-\frac12x}(1+O(x^{-1}))
\end{equation}
and
\begin{equation}\label{e-alphaasymp1}
  \alpha(x) = \int_{\infty}^x 2 (L_{N-1}^{(1)}(y))^2 e^{-y}dy
+ O(e^{-\frac32(1-\epsilon)x})
  = \frac{-2x^{2N-2}}{((N-1)!)^2}e^{-x}(1+O(x^{-1})),
\end{equation}
where $L^{(1)}_{N-1}(x)$ is the Laguerre polynomial 
of degree $N-1$ with parameter $1$ ;
(see, e.g. \cite{AS})
\begin{equation}
  L^{(1)}_{N-1}(x)=\sum_{j=0}^{N-1} \binom{N}{j+1}\frac{(-x)^{j}}{j!}.
\end{equation}
\item[(iii)] (Lax pair for Painlev\'e V)
The functions $\aLEN(x,\rho)$ and $\bLEN(x,\rho)$ are real and smooth 
in $x>0$ and $\rho>0$, 
and they satisfy the differential equations (note $w=\frac2{\rho}-1$)
\begin{equation}
  \frac{\partial}{\partial x} \begin{pmatrix} \bLEN \\ \aLEN \end{pmatrix} =
  -\frac12w\binom{\bLEN}{0} + \frac12 \beta \binom{\aLEN}{\bLEN},
\end{equation}
and
\begin{equation}
  \frac{\partial}{\partial w} \binom{\bLEN}{\aLEN} = -\frac12x \binom{\bLEN}{0} +
  \frac1{1-w^2} \begin{pmatrix} -(x\alpha)' & (x\beta)' \\ -(x\beta)' & (x\alpha)'
  \end{pmatrix}\binom{\bLEN}{\aLEN} -\frac12\frac{w}{1-w^2}x\beta \binom{\aLEN}{\bLEN}.
\end{equation}
\item[(iv)] (Asymptotics of $\aLEN, \bLEN$ as $x\to\infty$)
For any $0<\epsilon< 1/4$, as $x\to\infty$, we have
\begin{equation}\label{e-aasymp}
  \aLEN(x,\rho) =  \begin{cases}
1+ O(e^{-(1-2\epsilon)x}), &\qquad 0<\rho\le \frac{2}{2-\epsilon}, \\
\Phi(w;x) ( \Lambda_1(-w,x) + O(e^{-(1-2\epsilon)x}) ), &\qquad \rho> \frac{2}{2-\epsilon},
\end{cases}
\end{equation}
\begin{equation}\label{e-basymp}
  \bLEN(x,\rho) =  \begin{cases}
-\Lambda_1(w,x)+ O(e^{-(1-2\epsilon)x}\Phi(w;x)),
&\qquad 0<\rho\le \frac{2}{\epsilon}, \\
-\Phi(w;x) ( 1 + O(e^{-(1-2\epsilon)x}),
&\qquad \rho> \frac{2}{\epsilon},
\end{cases}
\end{equation}
where
\begin{equation}
  \Lambda_1(w,x,N) := \frac1{2 \pi i} \int_{|s-1|=\frac12|1-w|}
  \Phi(s;x) \frac{ds}{s-w},
\qquad \Phi(w;x) := e^{-\frac12xw}\bigl( \frac{1+w}{1-w} \bigr)^N.
\end{equation}
\item[(v)] (Asymptotics of $\aLEN, \bLEN$ as $\rho\to 0^+,2, \infty$)
We have
\begin{equation}\label{e-abatrho0}
  \lim_{\rho\to 0^+} \aLEN(x,\rho)= 1, \qquad
    \lim_{\rho\to 0^+} \bLEN(x,\rho)= 0
\end{equation}
and
\begin{equation}\label{e-abatrho2}
  \aLEN(x,2)= \ELEN(x)^{2}, \qquad
  \bLEN(x,2)= - \ELEN(x)^{2}.
\end{equation}
Also for fixed $y>0$,
\begin{equation}\label{e-abatrhoinfty}
  \lim_{\rho\to \infty} \aLEN(y\rho, \rho) = P(N,y), \qquad
\lim_{\rho\to\infty} \bLEN(y\rho,\rho) = 0,
\end{equation}
where $P(N,y)$ is the incomplete Gamma function (see, e.g. \cite{AS})
\begin{equation}\label{e-incompleteGamma}
  P(N,y) = \frac1{(N-1)!} \int_0^y e^{-t}t^{N-1} dt.
\end{equation}
\end{itemize}
\end{thm}

\begin{rem}
The existence of the solution $\theta$ (hence $\alpha$) to the 
Painlev\'e V equation \eqref{e-painV} with the asymptotic 
condition \eqref{e-alphaasymp1} is a part of the Theorem.
But as yet we do not have uniqueness for $\theta$ (or $\alpha$) 
as a solution of \eqref{e-painV} with 
asymptotic condition \eqref{e-alphaasymp1}.
Another missing piece of information is the asymptotics of $\alpha$ and $\beta$ 
as $x\to 0^+$. 
These issues will be studied in a later publication.
\end{rem}

From Proposition \ref{prop-inter}, 
by using \eqref{e-abatrho2} and \eqref{e-abatrho0}, Theorem
\ref{thm-PV} implies the following results for LOE and LSE 
at the special values $\rho=2$ ($A=0$) and $\rho\to 0^+$ ($A\to\infty$).

\begin{cor}\label{cor-LOE}
We have, with the notation 
$\R^N_{ord}(\xi):= \{\xi_N\le \xi_{N-1}\le\cdots \le \xi_1\}$,
for any $x >0$,
\begin{equation}\label{e-loe}
  \frac1{2^N\prod_{j=0}^{N-1}j!} \int_{\R^N_{ord}(\xi) \cap \{ \xi_1\le x\}} \Delta_N(\xi)
  \prod_{j=1}^N e^{-\frac12\xi_j}d\xi_j
  = \Prob( H^\symmO(N;2) \le x)=
  \ELEN(x)\FLEN(x),
\end{equation}
and for $N$ even,
\begin{equation}\label{e-lse}
  \frac1{\prod_{j=0}^{N-1}j!} \int_{\R^{N/2}_{ord}(\zeta) \cap \{ \zeta_1\le x\}}
  (\Delta_{N/2}(\zeta))^4
  \prod_{j=1}^{N/2} e^{-\zeta_j}d\zeta_j
  = \Prob( H^\symmO(N;0) \le x)=
  \frac12\biggl\{\bigl(\ELEN(x)\bigr)^{-1} + \ELEN(x) \biggr\}\FLEN(x).
\end{equation}
\end{cor}

\begin{rem}
Once the uniqueness of $\theta$ (or $\alpha$) 
is proven 
(see Remark above), and also 
the uniqueness of $\beta$ is established,
the above Corollary provides a tool for numerical computations of 
the largest eigenvalue distribution of LOE and LSE for the special 
case $w(x)=e^{-\frac12x}$ (see \eqref{e-Ldensity2}) for any finite $N$.
\end{rem}

For LUE, the largest eigenvalue distribution was obtained by
Tracy and Widom (\cite{TWFred}) :
\begin{equation}\label{e-lue}
  \frac1{\prod_{j=0}^{N-1}(j!)^2} \int_{\R^N_{ord}(\xi) \cap \{ \xi_1\le x\}} (\Delta_N(\xi))^2
  \prod_{j=1}^N e^{-\frac12\xi_j}d\xi_j
  = (\FLEN(x))^2.
\end{equation}
Note the special structure of the formulas 
\eqref{e-loe}, \eqref{e-lse} and \eqref{e-lue}, 
from which an interesting inter-relationship of the largest eigenvalues 
of LOE, LUE and LSE can be derived.
We refer the reader to \cite{ForRai99} for a 
full discussion on inter-relationship between orthogonal, 
symplectic and unitary ensembles.
We also remark
that Corollary \ref{cor-LOE} 
applies only for the case $\alpha=0$ of the Laguerre weight
\eqref{e-Ldensity2}. 
We do not have results for LOE and LSE of other values of $\alpha$.
On the other hand, 
LUE with different values of $\alpha\neq 0$ was analyzed in \cite{TWFred}.

If we take the limit $\rho\to\infty$ 
($A\downarrow -1/2$ ; note that \eqref{e-density} is defined for $A>-1/2$), 
we have
$\Prob(H^\symmO(N;\rho) \le x)\to 0$ 
for fixed $x>0$. 
To obtain a non-trivial limit from
\eqref{e-abatrhoinfty} we set $x=y\rho$ and 
let $\rho\to \infty$ while $y>0$ is fixed. 
Note that from \eqref{e-betaasymp1} and \eqref{e-alphaasymp1}, we have
$\FLEN(x)= 1- \frac{1}{2((N-1)!)} x^{2N-2}e^{-x} (1+O(x^{-1}))$ 
and $\ELEN(x)= 1- \frac{1}{(N-1)!} x^{N-1}e^{-\frac12 x} (1+O(x^{-1}))$ 
as $x\to\infty$, and hence we find that
$\ELEN(x)\FLEN(x)\to 1$ and $\ELEN(x)^{-1}\FLEN(x)\to 1$ as $x\to\infty$.

\begin{cor}\label{cor-rhobig}
For any fixed $y>0$,
\begin{equation}
  \lim_{\rho\to\infty} \Prob( H^\symmO(N;\rho) \le y\rho)
= P(N,y),
\end{equation}
where $P(N,y)$ is the incomplete Gamma function \eqref{e-incompleteGamma}.
\end{cor}

This result is consistent with the intuition that when $\rho$ is large, the longest
up/right path in the percolation model is basically the diagonal line through the points
$(i,i)$, $i=1,\cdots, N$. 
As $\rho\to\infty$, we expect that the random variable
$H^\symmO(N; \rho)$ is close to $S_N(\rho):= u(1,1)+u(2,2) +\cdots + u(N,N)$, the sum of
$N$ i.i.d. exponential random variables of mean $\rho$. 
A direct calculation shows that 
$\Prob(S_N(\rho) \le a)= P(N, a/\rho)$. Thus Corollary \ref{cor-rhobig} 
shows that indeed $H^\symmO(N;\rho) \sim S_N(\rho)$ when $\rho\to\infty$.

\bigskip

This paper is organized as follows. The proof of Proposition \ref{prop-inter} is given in
Section \ref{sec-inter}. In Section \ref{sec-conti}, 
we consider a different percolation
model which has 
a geometric random variable at each site instead of an exponential random
variable. Since a geometric random variable converges to an exponential 
random variable in an appropriate limit, 
the related percolation model with geometric random 
variables also
converges to the percolation model with exponential random variables (see Lemma
\ref{lem-Hlimit} below). 
In Section \ref{sec-conti} and Section \ref{sec-disc}, 
we present two different formulas for the geometric percolation model.
A multi-sum formula is given in Section \ref{sec-conti}, 
and in an appropriate limit, 
this multi-sum formula becomes the multi-integral formula in
Proposition \ref{prop-conti} for the exponential percolation model.
In Section \ref{sec-disc}, we express the distribution 
for the geometric percolation model in terms of  
orthogonal polynomials (see Lemma \ref{lem-disc3} below). 
The appropriate limit of these orthogonal polynomials yields 
the Painlev\'e expressions, Theorem \ref{thm-PV}. 
The asymptotics of orthogonal polynomials are obtained 
in Section \ref{sec-rhp1} by applying a steepest-descent method 
to the associated Riemann-Hilbert problem, 
and the proof of Theorem \ref{thm-PV} is given at the end of 
Section \ref{sec-rhp1}.
Some properties of the Riemann-Hilbert problem for 
the Painlev\'e V equation, which will be used in Section \ref{sec-rhp1} 
are summarized in Section \ref{sec-PV}.
Section \ref{sec-limit} presents some results for the percolation model 
as $N\to\infty$.
Finally, in Section
\ref{sec-corr} we show that the computation of 
Tracy and Widom (\cite{TracyWidomcluster}) which expresses 
the correlation functions for orthogonal and symplectic ensembles 
in terms of
determinants can also be applied to the density \eqref{e-dens} 
with $\beta=1$ which includes \eqref{e-density} as a special case.

\medskip
\noindent {\bf Acknowledgments.} 
We would like to thank Percy Deift, Eric Rains, Peter Forrester and Xin
Zhou for useful discussions.
Special thanks is due to Peter Forrester 
for keeping us informed of his recent work  
\cite{ForresterWPVI} with Nick Witte and \cite{ForresterR2} with Eric Rains. 
This work was supported in part by NSF Grant \# DMS
97-29992.

\section{Proof of Proposition \ref{prop-inter}}\label{sec-inter}

The identity \eqref{e-inter0} is trivial. We prove \eqref{e-interinfty} in this section. We
basically prove that when $A\to\infty$, the function $p(\xi_1, \cdots, \xi_N; A;w)$ on
$\R^N_{ord}$ will concentrate on the subset satisfying $\xi_{2j-1}=\xi_{2j}$ for each $j$.
We present the proof only for the case when $\beta=1$. But it would be clear that the proof
for general $\beta>0$ will be the same. In the below, we omit any dependence on $\beta$.

We assume that $N$ is even and $A\ge 1$. Substitute $\xi_{2j-1}=\xi_{2j}+x_j$, $j=1,\cdots,
N/2$, and set $\zeta_j=\xi_{2j}$. We use $\zeta_1,\cdots, \zeta_{N/2}$ and $x_1,\cdots,
x_{N/2}$ as new variables. Then the region of integration $\{\xi_N\le \xi_{N-1}\le\cdots\le
\xi_1\}$ becomes the set $\{\zeta_{N/2} \le\cdots \le \zeta_1 \}\cup  \{ 0\le x_1\}\cup\{
0\le x_j\le \zeta_{j-1}-\zeta_j, j=2,\cdots, N/2 \}$. We will denote by $\R^{N/2}_+(\zeta)$
the set $\{ 0\le x_1\}\cup\{ 0\le x_j\le \zeta_{j-1}-\zeta_j, j=2,\cdots, N/2 \}$. In the
below, we use the notation $d\zeta=\prod_{j=1}^{N/2} d\zeta_j$ and $dx=\prod_{j=1}^{N/2}
dx_j$, and also $w(\zeta)=\prod_{j=1}^{N/2} w(\zeta_j)$. Then the integral on the left hand
side of \eqref{e-interinfty} without the limit $A\to +\infty$ becomes
\begin{equation}\label{e-intermain}
 (*):= Z_{N,A;w}^{-1} \int_{\R^{N/2}_{ord}} w(\zeta) d\zeta
  \int_{\R^{N/2}_+(\zeta)} f
  p(x,\zeta) \prod_{j=1}^{N/2} e^{-Ax_j} w(x_j+\zeta_j) dx_j,
\end{equation}
where
\begin{equation}\label{e-p}
\begin{split}
  p(x,\zeta)
&= \prod_{i=1}^{N/2} x_i \prod_{1\le i< j\le N/2} (x_i-x_j+\zeta_i-\zeta_j) \\
&\times  \prod_{1\le i<j\le N/2} (x_i+\zeta_i-\zeta_j)
\prod_{1\le i<j\le N/2} (\zeta_i-\zeta_j-x_j)
  \prod_{1\le i<j\le N/2} (\zeta_i-\zeta_j)
\end{split}
\end{equation}
is a positive polynomial, and also each factor is non-negative.

Let $0<\epsilon<1$ be any fixed number. Let $B^{N/2}_\epsilon := \{ 0\le x_j\le\epsilon,
j=1,\cdots, N/2 \}$. We divide the integral with respect to $x$ into two regions : (1)
$X_1^\epsilon:=\R^{N/2}_+(\zeta)\cap B^{N/2}_\epsilon$ and (2) $X_2^\epsilon:=
\R^{N/2}_+(\zeta)\setminus X_1^\epsilon$.

We first show that $Z_{N,A;w}^{-1}$ has a polynomial growth as $A\to\infty$.
\begin{lem}\label{lem-inter}
We have
\begin{equation}
  \lim_{A\to\infty} A^{N} Z_{N,A;w} = Z_{N,\infty; w}.
\end{equation}
where $Z_{N,\infty; w}$ is defined in \eqref{e-Zinfty}.
\end{lem}

\begin{proof}
For $x\in X_2^\epsilon$, since $e^{-Ax_j}\le 1$ and
there is at least one $x_j$ satisfying $x_j>\epsilon$, we have
\begin{equation}\label{e-inter1}
\begin{split}
  &\int_{\R^{N/2}_{ord}} w(\zeta)d\zeta \int_{X_2^\epsilon} p(x,\zeta) \prod_{j=1}^{N/2}
  e^{-Ax_j} w(x_j+\zeta_j) dx_j \\
  &\qquad \le e^{-A\epsilon} \int_{\R^{N/2}_{ord}} w(\zeta)d\zeta \int_{X_2^\epsilon} p(x,\zeta)
  \prod_{j=1}^{N/2} w(x_j+\zeta_j) dx_j
  \le e^{-A\epsilon} Z_{N,0;w}.
\end{split}
\end{equation}
For $x\in X_1^\epsilon$, since $\|V'\|_{L^\infty(\R)} \le C_0$,
we have $w(\zeta_j) e^{-\epsilon C_0}\le w(x_j+\zeta_j)
\le w(\zeta_j) e^{\epsilon C_0}$
for all $1\le j\le N/2$.
Thus
\begin{equation}\label{e-bound1}
  Z_{N,A;w} \le  e^{\frac12\epsilon NC_0}\int_{\R^{N/2}_{ord}}
\prod_{j=1}^{N/2} w(\zeta_j)^2d\zeta_j
\int_{X_1^\epsilon} p(x,\zeta) \prod_{j=1}^{N/2} e^{-Ax_j} dx_j
+ e^{-A\epsilon} Z_{N,0;w},
\end{equation}
and
\begin{equation}\label{e-bound2}
  Z_{N,A;w} \ge  e^{-\frac12\epsilon NC_0}\int_{\R^{N/2}_{ord}}
\prod_{j=1}^{N/2} w(\zeta_j)^2d\zeta_j
\int_{X_1^\epsilon} p(x,\zeta) \prod_{j=1}^{N/2} e^{-Ax_j} dx_j.
\end{equation}

In \eqref{e-p}, for $x\in X_1^\epsilon$ and $\zeta\in \R^{N/2}_{ord}$, we have
\begin{equation}\label{e-pless}
\begin{split}
  p(x,\zeta) & \le
\prod_{i=1}^{N/2} x_i
\prod_{i<j} (\epsilon + \zeta_i-\zeta_j)
\prod_{i<j} (\epsilon + \zeta_i-\zeta_j) \prod_{i<j} (\zeta_i-\zeta_j)
\prod_{i<j} (\zeta_i-\zeta_j) \\
& \le (\Delta_{N/2}(\zeta)^4+\epsilon Q(\zeta)) \prod_{k=1}^{N/2} x_k,
\end{split}
\end{equation}
where $\Delta$ is the Vandermonde product \eqref{e-Delta},
and $Q(\zeta)$ is a positive polynomial. Thus using
\begin{equation}\label{e-basic}
\int_0^\infty xe^{-Ax}dx = A^{-2},
\end{equation}
\eqref{e-bound1} is less than or equal to
\begin{equation}
  e^{\frac12 \epsilon NC_0}\frac{1}{A^{N}} \int_{\R^{N/2}_{ord}}
(\Delta_{N/2}(\zeta)^4+\epsilon Q(\zeta))
\prod_{j=1}^{N/2} w(\zeta_j)^2 d\zeta_j
+ e^{-A\epsilon} Z_{N,0;w}.
\end{equation}
But
\begin{equation}
  \int_{\R^{N/2}_{ord}}
\Delta_{N/2}(\zeta)^4 \prod_{j=1}^{N/2}
w(\zeta_j)^2 d\zeta_j = Z_{N,\infty;w},
\end{equation}
and hence
\begin{equation}
  Z_{N,A;w} \le \frac{1}{A^{N}} e^{\frac12 \epsilon NC_0}
( Z_{N,\infty;w}+\epsilon C_1)
+ e^{-A\epsilon} Z_{N, 0;\infty},
\end{equation}
for some constant $C_1>0$.
This implies
\begin{equation}
  \limsup_{A\to\infty} A^{N}  Z_{N,A;w} \le Z_{N,\infty;w}.
\end{equation}

Similarly to \eqref{e-pless}, we have
\begin{equation}
  p(x,\zeta) \ge (\Delta_{N/2}(\zeta)^4-\epsilon R(\zeta))
\prod_{k=1}^{N/2} x_k
\end{equation}
for some positive polynomial $R(\zeta)$.
Hence from \eqref{e-bound2},
\begin{equation}\label{e-lower2}
  Z_{N,A;w}\ge  e^{-\frac12 \epsilon NC_0} \int_{\R^{N/2}_{ord}}
(\Delta_{N/2}(\zeta)^4-\epsilon R(\zeta))
\prod_{j=1}^{N/2} w(\zeta_j)^2 d\zeta_j
\int_{X_1^\epsilon} \prod_{j=1}^{N/2} x_j e^{-Ax_j} dx_j.
\end{equation}
Since $X_1^\epsilon= \R^{N/2}_+(\zeta) \cap B^{N/2}_\epsilon$,
we have
\begin{equation}\label{e-lower1}
  \int_{X_1^\epsilon} \prod_{j=1}^{N/2} x_j e^{-Ax_j} dx_j
= \frac1{A^{N}}\prod_{j=1}^{N/2}
(1-(A+1)e^{-A \min (\epsilon, \zeta_{j-1}-\zeta_j)}),
\end{equation}
where $\zeta_0:= +\infty$.
We note that \eqref{e-lower1} is bounded below
by $A^{-N} (1-(A+1)e^{-A\epsilon})^N$, and above by $A^{-N}$,
if we take $A$ large enough.
Hence by considering \eqref{e-lower2} as sum of two integrals,
one involving $\Delta_{N/2}^4$
and the other involving $R$, we find that
\begin{equation}
  Z_{N,A;w} \ge e^{-\frac12 \epsilon NC_0} \frac{Z_{N,\infty;w}}{A^{N}}
(1-(A+1)e^{-A\epsilon})^N
- \epsilon e^{-\frac12 \epsilon NC_0} \frac{C_2}{A^{N}}
\end{equation}
for some constant $C_2>0$.
Therefore we obtain
\begin{equation}\label{e-Zless}
  Z_{N,\infty;w} \le \liminf_{A\to\infty} A^{N} Z_{N,A;w}.
\end{equation}

\end{proof}

\subsection*{Proof of \eqref{e-interinfty}}

Fix $0<\epsilon<1$.
Then there is a $\delta>0$ such that for $x\in X_1^{\delta}$, we have
\begin{equation}\label{e-delta}
  \bigl| f(\zeta_1+x_1,\zeta_1,\zeta_2+x_2,\zeta_2,\cdots) -
g(\zeta_1,\zeta_2, \cdots)\bigr| < \epsilon.
\end{equation}
We write \eqref{e-intermain} as
\begin{equation}\label{e-interdiff}
  (*) = (**)+ (*1)+(*2)+(*3),
\end{equation}
where
\begin{equation}\label{e-2star}
  (**) = Z_{N,A;w}^{-1} \int_{\R^{N/2}_{ord}} g w(\zeta)d\zeta
  \int_{\R^{N/2}_+(\zeta)}
  p(x,\zeta) \prod_{j=1}^{N/2} e^{-Ax_j} w(x_j+\zeta_j) dx_j,
\end{equation}
and
\begin{eqnarray}
 (*1)&=& Z_{N,A;w}^{-1}\int_{\R^{N/2}_{ord}}  w(\zeta)d\zeta
  \int_{X_2^{\delta}} f
  p(x,\zeta) \prod_{j=1}^{N/2} e^{-Ax_j} w(x_j+\zeta_j) dx_j, \\
  (*2) &=& - Z_{N,A;w}^{-1} \int_{\R^{N/2}_{ord}}  g w(\zeta)d\zeta
  \int_{X_2^{\delta}}
  p(x,\zeta) \prod_{j=1}^{N/2} e^{-Ax_j} w(x_j+\zeta_j) dx_j, \\
  (*3) &=& Z_{N,A;w}^{-1} \int_{\R^{N/2}_{ord}}  w(\zeta)d\zeta
  \int_{X_1^{\delta}} (f-g)
  p(x,\zeta) \prod_{j=1}^{N/2} e^{-Ax_j} w(x_j+\zeta_j) dx_j.
\end{eqnarray}
As in \eqref{e-inter1}, we have
\begin{equation}
  |(*1)| \le  \| f\|_{L^\infty} e^{-A\delta} \frac{Z_{N,0;w}}{Z_{N,A;w}},
\qquad
  |(*2)| \le \| g\|_{L^\infty} e^{-A\delta} \frac{Z_{N,0;w}}{Z_{N,A;w}}.
\end{equation}
On the other hand, from \eqref{e-delta},
\begin{equation}
  |(*3)| \le \epsilon Z_{N,A;w}^{-1} \int_{\R^{N/2}_{ord}}  w(\zeta)d\zeta
  \int_{X_1^{\delta}}
p(x,\zeta) \prod_{j=1}^{N/2} e^{-Ax_j} w(x_j+\zeta_j) dx_j
\le \epsilon.
\end{equation}
Thus using Lemma \ref{lem-inter},
$\limsup_{A\to\infty} |(*1)+(*2)+(*3)| \le \epsilon$,
but $\epsilon>0$ is arbitrarily small,
hence we have
\begin{equation}
  \lim_{A\to\infty} |(*)-(**)| =0.
\end{equation}

Now the only remaining thing is to show that as $A\to\infty$.
\begin{equation}\label{e-only}
\begin{split}
   (**) \to
  Z_{N,\infty; w}^{-1} \int_{\R^{N/2}_{ord}} g
   \prod_{1\le i<j\le N/2}(\zeta_i-\zeta_j)^4
   \prod_{j=1}^{N/2} (w(\zeta_j))^2 d\zeta_j.
\end{split}
\end{equation}
Fix $0<\epsilon'<1$.
We write
\begin{equation}
  (**)= (**1)+(**2)+(**3),
\end{equation}
where
\begin{eqnarray}
  (**1)&=& Z_{N,A;w}^{-1}\int_{\R^{N/2}_{ord}}  g w(\zeta)d\zeta
  \int_{X_2^{\epsilon'}}
  p(x,\zeta) \prod_{j=1}^{N/2} e^{-Ax_j} w(x_j+\zeta_j) dx_j, \\
  (**2) &=& Z_{N,A;w}^{-1} \int_{\R^{N/2}_{ord}}  g w(\zeta)d\zeta
  \int_{X_1^{\epsilon'}}
  p(x,\zeta) \prod_{j=1}^{N/2} e^{-Ax_j} (w(x_j+\zeta_j)-w(\zeta_j)) dx_j, \\
  (**3) &=& Z_{N,A;w}^{-1} \int_{\R^{N/2}_{ord}}  g w(\zeta)^2d\zeta
  \int_{X_1^{\epsilon'}}
  p(x,\zeta) \prod_{j=1}^{N/2} e^{-Ax_j} dx_j.
\end{eqnarray}
As in \eqref{e-inter1}, we find
\begin{equation}\label{e-2star1}
  |(**1)| \le \|g\|_{L^\infty} Z_{N,A;w}^{-1} e^{-A\epsilon'} Z_{N,0;w}
\to 0,
\end{equation}
as $A\to\infty$, using Lemma \ref{lem-inter}.
For the estimation of $(**2)$, we use
$|w(x_j+\zeta_j) - w(x_j)| \le D(\epsilon') w(x_j+\zeta_j)$
for $x_j\in X_1^{\epsilon'}$,
where $D(\epsilon'):=\max (|1-e^{\epsilon' C_0}|, |1-e^{-\epsilon' C_0}|)$,
and find that
\begin{equation}\label{e-2star2}
\begin{split}
  |(**2)| \le D(\epsilon') \|g\|_{L^\infty}
Z_{N,A;w}^{-1} \int_{\R^{N/2}_{ord}}   w(\zeta)d\zeta
  \int_{X_1^{\epsilon'}}
  p(x,\zeta) \prod_{j=1}^{N/2} e^{-Ax_j} w(x_j+\zeta_j) dx_j
\le D(\epsilon') \|g\|_{L^\infty}
\end{split}
\end{equation}
where the second inequality is obtained by replacing the region
$X_1^{\epsilon'}$ by $\R^{N/2}_+(\zeta)$ and noting the
the total integral is $1$ by \eqref{e-intermain}.
For $(**3)$, we note that \eqref{e-pless}, \eqref{e-basic} and
Lemma \ref{lem-inter}
yield
\begin{equation}
\begin{split}
  &0\le Z_{N,A;w}^{-1} \int_{X_1^{\epsilon'}}
  p(x,\zeta) \prod_{j=1}^{N/2} e^{-Ax_j} dx_j
\le (\Delta_{N/2}(\zeta)^4+\epsilon'Q(\zeta)) Z_{N,A;w}^{-1}
\int_{X_1^{\epsilon'}} \prod_{j=1}^{N/2} x_je^{-Ax_j} dx_j \\
&\qquad \le (\Delta_{N/4}(\zeta)^4+\epsilon'Q(\zeta)) Z_{N,A;w}^{-1} A^{-N}
\le (\Delta_{N/4}(\zeta)^4+\epsilon'Q(\zeta))c Z_{N,\infty;w}^{-1}.
\end{split}
\end{equation}
for some constant $c>1$. But $\Delta_{N/4}(\zeta)^4+\epsilon'Q(\zeta)$ is integrable with
respect to the measure $w(\zeta)^2d\zeta$ on $\R^{N/2}_{ord}$, and hence we can use the
Lebesgue dominated convergence theorem. But an argument similar to
\eqref{e-pless}-\eqref{e-Zless} shows that for each $\zeta\in \R^{N/2}_{ord}$,
\begin{equation}
  \lim_{A\to\infty}  Z_{N,A;w}^{-1} \int_{X_1^{\epsilon'}}
p(x,\zeta) \prod_{j=1}^{N/2} e^{-Ax_j} dx_j
= Z_{N,\infty;w}^{-1} \Delta_{N/4}(\zeta)^4,
\end{equation}
and we find
\begin{equation}\label{e-2star3}
  \lim_{A\to\infty} (**3) =
  Z_{N,\infty; w}^{-1} \int_{\R^{N/2}_{ord}} g
   \prod_{1\le i<j\le N/2}(\zeta_i-\zeta_j)^4
   \prod_{j=1}^{N/2} (w(\zeta_j))^2 d\zeta_j.
\end{equation}
Thus the estimates \eqref{e-2star1}, \eqref{e-2star2}, \eqref{e-2star3}, together with the
fact that $D(\epsilon')\to 0$ as $\epsilon\to 0$, yield \eqref{e-only}, and we prove the
Proposition \ref{prop-inter}.

\subsection*{Proof of \eqref{e-char}}

In the analysis above, the fact that $f$ is uniformly continuous is used
only for the estimation of $(*3)$.
Now when $f= 1_{\xi_N\le \cdots\le \xi_1\le t}
= 1_{0\le x_1\le t-\zeta_1,\zeta_1<t}$ and
$g=1_{\zeta_1\le t}$,
$(*3)$ becomes
\begin{equation}
\begin{split}
  (*3) =
&Z_{N,A;w}^{-1} \int_{\R^{N/2}_{ord}\cap \{ \zeta_1<t\}}
w(\zeta)d\zeta
  \int_{X_1^{\delta}\cap \{0\le x_1\le t-\zeta_1\}}
  p(x,\zeta) \prod_{j=1}^{N/2} e^{-Ax_j} w(x_j+\zeta_j) dx_j \\
& - Z_{N,A;w}^{-1} \int_{\R^{N/2}_{ord}\cap \{ \zeta_1<t\}}
w(\zeta)d\zeta
  \int_{X_1^{\delta} }
  p(x,\zeta) \prod_{j=1}^{N/2} e^{-Ax_j} w(x_j+\zeta_j) dx_j
\end{split}
\end{equation}
where now $0<\delta<1$ can be taken to be an arbitrary fixed constant.
When $t-\zeta_1>\delta$, $X_1^{\delta}\cap \{0\le x_1\le t-\zeta_1\}
= X_1^{\delta}$, and the above two integrals in $x$ are the same,
and hence we have
\begin{equation}
  |(*3)| \le
Z_{N,A;w}^{-1} \int_{\R^{N/2}_{ord}\cap \{ t-\delta\le \zeta_1<t\}}
w(\zeta)d\zeta
  \int_{X_1^{\delta}\cap \{t-\zeta_1<x_1\le \delta \}}
  p(x,\zeta) \prod_{j=1}^{N/2} e^{-Ax_j} w(x_j+\zeta_j) dx_j.
\end{equation}
As in Lemma \ref{lem-inter}, this can be estimated as
\begin{equation}
  |(*3)| \le Z_{N,A;w}^{-1} A^N e^{\frac12\delta N C_0}
\int_{\R^{N/2}_{ord}\cap \{ t-\delta\le \zeta_1<t\}} (\Delta_{N/4}(\zeta)^4+\delta
Q(\zeta)) w(\zeta)^2d\zeta.
\end{equation}
But $Z_{N,A;w}^{-1} A^N$ is bounded as $A\to\infty$, and the integral vanishes when we take
$\delta\to 0$, and hence we obtain $(*3)\to 0$ as $A\to\infty$. The rest of analysis is the
same as for the proof of \eqref{e-interinfty}.

\subsubsection*{Remark}

In addition to the change $\beta \to 4\beta$, we can also obtain
the transition $\beta \to k^2 \beta$ for each $k\in \N$.
Let
\begin{equation}
  \eta_3(\xi; A):= e^{-A(\xi_1+\xi_2-2\xi_3 + \xi_4+\xi_5-2\xi_6+\cdots)},
\end{equation}
and similarly we set $\eta_k(\xi; A)$ with the
term $\xi_1+\xi_2+\cdots +\xi_{k-1} - (k-1)\xi_k + \xi_{k+1}-\cdots$.
Then when $N$ is a multiple of $k$,
\begin{equation}\label{e-242}
\begin{split}
  & \lim_{A\to +\infty} Z_{N,A;w;\beta;k}^{-1}\int_{\R^N_{ord}} f(\xi)
  \Delta_N(\xi)^\beta \eta_k(\xi; A)
  \prod_{j=1}^N w(\xi_j) d\xi_j  \\
  &\qquad =
  Z_{N,\infty; w;\beta;k}^{-1} \int_{\R^{N/k}_{ord}} f_k(\zeta)
   \Delta_{N/k}(\zeta)^{k^2\beta}
   \prod_{j=1}^{N/k} (w(\zeta_j))^k d\zeta_j.
\end{split}
\end{equation}
where $f_k$ is obtained from $f$ by setting the first $k$ variables equal, and the next $k$
variables equal, and so on. There are other possible choices of $\eta_k$. For instance,
\begin{equation}
  \eta_k(\xi; A) = e^{-A(\xi_1-\xi_k+\xi_{k+1}-\xi_{2k}+\cdots)}
\end{equation}
would again yield \eqref{e-242}.

\section{First formula for geometric percolation : multi-sum expression}
\label{sec-conti}

For $0<q<1$, let $g(q)$ denote the geometric random variable with parameter $q$ : for
$k=0,1,2,\cdots$, $\Prob(g(q)=k)=(1-q)q^k$. We consider a last passage percolation model
with geometric random variables analogous to the percolation model with exponential random
variables considered in the Introduction. 
Namely, to each site $(i,j)\in \Z_+^2$ the
random variable $X(i,j)$ is attached where
\begin{align}
  X(i,j) &\sim g(q), \qquad i<j, \\
  X(i,i) &\sim g(\alpha \sqrt{q}), \\
  X(j,i)&=X(j,i), \qquad i<j.
\end{align}
Here $q\in(0,1)$ and $\alpha\in(0,1/\sqrt{q})$ are fixed numbers. As before, the random
variables are independent except the symmetry condition $X(i,j)=X(j,i)$. As in the
Introduction, $\Pi(N)$ denotes the set of up/right paths from $(1,1)$ to $(N,N)$. We define
\begin{equation}\label{e-Ggeom}
  G^\symmO(N;\alpha) = \max\bigl\{ \sum_{(i,j)\in\pi} X(i,j)
: \pi \in \Pi(N)\bigr\}.
\end{equation}

Since a proper limit of the geometric random variable becomes the exponential random
variable, we find that the random variable $H^\symmO(N;\rho)$ is a limit of
$G^\symmO(N.\alpha)$.

\begin{lem}\label{lem-Hlimit}
We have
\begin{equation}\label{e-GtoH}
  \Prob(H^\symmO(N;\rho)\le x) = \lim_{L\to\infty}
\Prob( G^\symmO(N;\alpha) \le xL),
\end{equation}
where we set
\begin{equation}
\label{e-contiL1}
  \sqrt{q} =1-\frac{1}{2L}, \qquad \alpha =1-\biggl(\frac1{\rho}-\frac12\biggr)\frac{1}{L}.
\end{equation}
\end{lem}

\begin{proof}
It is direct to check that with \eqref{e-contiL1},
\begin{equation}
  e(1)=\lim_{L\to\infty} \frac{g(q)}{L}, \qquad
  e(\rho)=\lim_{L\to\infty} \frac{g(\alpha\sqrt{q})}{L},
\end{equation}
in distribution.
Thus under this limit,
the last passage percolation model with geometric random variables
becomes the last passage percolation model with exponential random variables.
It is also direct to check that $G^\symmO(N,\alpha)/L \to H^\symmO(N,\rho)$
in distribution.
\end{proof}

Now the key thing is that there are two different formulas 
for $\Prob(G^\symmO(N,\alpha)\le n)$. 
Thus by taking the exponential limit $L\to\infty$ 
with \eqref{e-contiL1}, we would
obtain two different formulas for $\Prob(H^\symmO(N,\rho)\le x)$. 
It will turn out that 
one of the limiting formula is the multi-integral
formula given in the right-hand-side of \eqref{e-multiint} 
which represents the probability
distribution for the rightmost `particle' in the interpolating ensemble,
and the other is the Painlev\'e V 
expression, Theorem \ref{thm-PV}. 
We present the first formula 
for $\Prob(G^\symmO(N,\alpha)\le n)$ in this section. 
The second formula will be considered in the subsequent sections.

The following lemma is modeled on the paper \cite{kurtj:shape} 
in which a similar
result for the unsymmetrized case 
(at each site $(i,j)$, the geometric random variables
$X(i,j)$ are independent and identically distributed 
without the symmetry condition
$X(i,j)=X(j,i)$).

\begin{lem}\label{lem-disc}
Let $\N_0=\N\cup\{0\}$.
We have
\begin{eqnarray}\label{e-disc}
  \Prob( G^\symmO(N;\alpha) \le n) = Z_1(N,\alpha)^{-1}
\sum_{\substack{0\le h_N< \cdots<h_1\le n+N-1 \\ h_j\in \N_0}}
\prod_{1\le i<j\le N} (h_i-h_j) \prod_{i=1}^N q^{h_i/2}\alpha^{-(-1)^jh_j},
\end{eqnarray}
where the normalization constant is
\begin{equation}\label{e-discZ}
  Z_1(N,\alpha)=(1-\alpha\sqrt{q})^{-N}(1-q)^{-N(N-1)/2}\alpha^{[N/2]}
q^{N(N-1)/2} \prod_{j=0}^{N-1} j!.
\end{equation}
\end{lem}

\begin{rem}
The result for the special case when $\alpha=1$ is stated in
Remark 5.2 \cite{kurtj:shape}.
\end{rem}

\begin{proof}(cf. Section 2.1, \cite{kurtj:shape} and
Proof of Theorem 7.1, \cite{BR1})
Let $\lambda=(\lambda_1,\cdots,\lambda_N)$ is a Young diagram
: a sequence of integers, $\lambda_1\ge\cdots\ge\lambda_N\ge 0$.
The number $d_\lambda(N)$ of semistandard Young tableaux (SSYT) of shape
$\lambda$ with elements taken from $\{1,\cdots,N\}$ is equal to
(see e.g., \cite{Stl})
\begin{equation}\label{e-SSYT}
  d_\lambda(N)=\prod_{1\le i<j\le N} \frac{\lambda_i-\lambda_j+j-i}{j-i}
= \prod_{j=0}^{N-1} \frac1{j!}\prod_{1\le i<j\le N} (h_i-h_j),
\qquad h_j:=\lambda_j+N-j.
\end{equation}
Let $D(N)=\sum_{j=1}^N X(j,j)$ and $OD(N)=\sum_{1\le i<j\le N} X(i,j)$.
Then
\begin{equation}\label{e-discpf1}
\begin{split}
  &\Prob( G^\symmO(N;\alpha) \le n) \\
&= \sum_{k,m\ge 0}
\Prob(G^\symmO(N;\alpha) \le n | D(N)=m, OD(N)=k ) \Prob(D(N)=m, OD(N)=k )
\end{split}
\end{equation}
Let $I^{m,k}_N$ be the set of $N\times N$ symmetric matrices
with non-negative integer-valued entries
such that the sum of diagonal entries is $m$
and the sum of upper-triangular entries is equal to $k$.
For $A=(a_{ij})_{1\le i,j\le N}\in I^{m,k}_N$,
the probability that the random matrix
$(X(i,j))_{1\le i,j\le N}$ becomes $A$ is given by
\begin{equation}
  \prod_{j}(1-\alpha\sqrt{q})(\alpha\sqrt{q})^{a_{jj}}
\prod_{i<j}(1-q)q^{a_{ij}}
=(1-\alpha\sqrt{q})^N(1-q)^{N(N-1)/2}\alpha^mq^{(2k+m)/2},
\end{equation}
which is a value independent of the choice of $A\in I^{m,k}_N$.
Hence we have
\begin{equation}\label{e-discpf2}
\Prob(D(N)=m, OD(N)=k )
=\#I^{k,m}_N (1-\alpha\sqrt{q})^N(1-q)^{N(N-1)/2}\alpha^mq^{(2k+m)/2}.
\end{equation}
Also the conditional probability $\Prob(\cdot | D(N)=m, OD(N)=k )$
is the uniform distribution on $I^{k,m}_N$.
Now we use a version of the Robinson-Schensted-Knuth correspondence
which yields a bijection between $I^{k,m}_N$
and the set of semistandard Young tableaux with shapes
$\lambda\vdash 2k+m$ such that $f(\lambda)=m$,
where
\begin{equation}\label{e-flambda}
  f(\lambda)=\lambda_1-\lambda_2+\lambda_3-\lambda_4+\cdots
\end{equation}
denotes the number of odd columns of $\lambda$.
Moreover, under the Robinson-Schensted-Knuth correspondence,
the length of the longest up/right path $G^\symmO$ is equal to the
first row $\lambda_1$.
Therefore, we have
\begin{equation}\label{e-discpf3}
  \Prob(G^\symmO(N;\alpha) \le n | D(N)=m, OD(N)=k )
=\frac1{\#I^{k,m}_N} \sum_{\substack{\lambda\vdash 2k+m \\ f(\lambda)=m \\
\lambda_1\le n}} d_\lambda(N).
\end{equation}
Thus from \eqref{e-discpf1}, \eqref{e-discpf2} and \eqref{e-discpf3},
\begin{equation}\label{e-discpf4}
   \Prob( G^\symmO(N;\alpha) \le n) =
(1-\alpha\sqrt{q})^N(1-q)^{N(N-1)/2}\sum_{\lambda_1\le n}
d_\lambda(N) \alpha^{f(\lambda)}q^{(\lambda_1+\lambda_2+\cdots)/2}
\end{equation}
where the sum is over all Young diagrams $\lambda$ satisfying $\lambda_1\le n$.
Note that if $\lambda=(\lambda_1,\cdots\lambda_k)$ with $\lambda_k>0$
and $k>N$, we have $d_\lambda(N)=0$, and hence the sum in \eqref{e-discpf4}
is over the Young diagrams $\lambda=(\lambda_1,\cdots,\lambda_N)$
satisfying $\lambda_1\le n$.
Now we use \eqref{e-SSYT} and set $h_j=\lambda_n+N-j$,
and the results \eqref{e-disc} and \eqref{e-discZ} are obtained.
\end{proof}

By taking the limit \eqref{e-GtoH} of \eqref{e-disc}, we obtain Proposition
\ref{prop-conti}, which was originally given in Remark 7.6.1, \cite{BR1}. 

\section{Second formula for geometric percolation : orthogonal polynomials}\label{sec-disc}

In this section, we present the second formula for the geometric percolation model
introduced in Section \ref{sec-conti}. The following determinant formula is given in
Theorem 7.1 of \cite{BR1} for a more general model.

\begin{lem}\label{lem-disc2}
(Theorem 7.1, \cite{BR1})
We have
\begin{align}\label{e-GEO}
  \Prob( G^\symmO(N;\alpha) \le n) = Z_3(N,\alpha)^{-1} \Exp_{U\in O(n)}
\det((1+\alpha U)(1+\sqrt{q}U)^N)
\end{align}
with
\begin{equation}
  Z_3(N,\alpha) = (1-\alpha \sqrt{q})^{-N}(1-q)^{-N(N-1)/2}.
\end{equation}
\end{lem}

\begin{proof}
The random variable $G^\symmO(N;\alpha)$ is same as
$\ell^\symmO_{W}(\overrightarrow{q};\alpha)$ in Theorem 7.1 of \cite{BR1},
where $W=\Z_+$ and $\overrightarrow{q}=(\sqrt{q},\cdots, \sqrt{q}, 0,0,\cdots)$
with $\sqrt{q}$'s occurring $N$ times. (In Section 7 of \cite{BR1},
$\overrightarrow{q}$ is denoted by $q$, but we take this notation to avoid
the confusion with the parameter $0<q<1$ of the geometric random variable.)
By (7.32) of \cite{BR1}, we have
\begin{equation}\label{e-disc1}
  \Prob(  G^\symmO(N;\alpha) \le n) =
\Prob(  \ell^\symmO_{\Z_+}(\overrightarrow{q};\alpha) \le n)
= Z^\symmO_{\Z_+}(\overrightarrow{q};\alpha) \sum_{\ell(\lambda)\le n}
\alpha^{f(\lambda)}s_{\lambda}(/{\overrightarrow{q}}_{\Z_+})
\end{equation}
where $\ell(\lambda)$ is the number of parts of the Young diagram $\lambda$
and $f(\lambda)$ is same as \eqref{e-flambda}.
Indeed this formula is, after the modification $\lambda\mapsto \lambda^t$,
is equal to the formula \eqref{e-discpf4} in Section \ref{sec-conti}.
The normalization constant is by (7.11) of \cite{BR1},
\begin{equation}
  Z_3(N,\alpha) = Z^\symmO_{\Z_+}(\overrightarrow{q};\alpha)^{-1}
=(1-\alpha \sqrt{q})^{-N}(1-q)^{-N(N-1)/2}.
\end{equation}
From (5.55) of \cite{BR1} with the modification as in the paragraph
preceding Theorem 7.1 of \cite{BR1},
the sum in \eqref{e-disc1} is equal to
\begin{equation}
  \sum_{\ell(\lambda)\le n}
\alpha^{f(\lambda)}s_{\lambda}(/{\overrightarrow{q}}_{\Z_+})
= \Exp_{U\in O(n)} \det((1+\alpha U)
H(U;\overrightarrow{0}/{\overrightarrow{q}}_{\Z_+})).
\end{equation}
But by (7.28) and (5.6) of \cite{BR1}, we have
\begin{equation}
  H(U;\overrightarrow{0}/{\overrightarrow{q}}_{\Z_+})
= E(U; {\overrightarrow{q}}_{\Z_+})
= (1+\sqrt{q}U)^N,
\end{equation}
and hence we obtain \eqref{e-GEO}.
\end{proof}

Let $O(n)_{\pm}$ denote the connected component of $O(n)$ with $\det(U)=\pm 1$,
respectively. The authors in \cite{BR1} expressed the expected value over the orthogonal
group in \eqref{e-GEO} in terms of the related orthogonal polynomials. Set
\begin{equation}
  \psi(z)= \psi(z;M,N):=(1+\sqrt{q}z)^N(1+\sqrt{q}z^{-1})^N.
\end{equation}
Let $\pi_j(z)$ be the monic orthogonal polynomial of degree $j$
with respect to the measure $\psi(z)dz/(2\pi iz)$
on the unit circle :
\begin{equation}\label{e-firstop}
  \int_{|z|=1} \pi_j(z)\overline{\pi_k(z)} \psi(z)\frac{dz}{2\pi iz}
= N_j \delta_{jk},
\end{equation}
where $N_j$ is the square of the norm of $\pi_j(z)$. We also set
\begin{equation}\label{}
  \pi^*_j = z^j \pi_j(z^{-1}).
\end{equation}
\begin{rem}
  In general, $\pi_j^*$ is defined by $z^j\overline{\pi_j}(z^{-1})$. But for the case at
  hand, all the coefficients of $\pi_j$ are real, and hence taking the complex conjugate
  has no effect.
\end{rem}
We also set
\begin{eqnarray}\label{e-phidefine}
  \phi_j(z) &= (1+\sqrt{q}z)^{N}\pi_j(z),\\
\label{e-phistardefine}
  \phi^*_j(z) &= (1+\sqrt{q}z)^N \pi^*(z).
\end{eqnarray}
Note that a version of strong Szeg\"o theorem yields that (see e.g., \cite{Jo1})
\begin{equation}
  \lim_{n\to\infty} \Exp_{U\in O(n)_{\pm}}
\det((1+\sqrt{q}U)^N) = (1-q)^{-N(N-1)/2}.
\end{equation}
From Theorem 3.1 and Theorem 2.3 in \cite{BR1}, \eqref{e-GEO} has the following expression.

\begin{lem}\label{lem-disc3}
For $n\ge 1$,
\begin{align}
  \Prob( G^\symmO(N;\alpha) \le 2n) &=
\frac12\biggl\{ [\phi^*_{2n-1}(-\alpha)-\alpha\phi_{2n-1}(-\alpha)]
\Delta^{--}_n
+[\phi^*_{2n-1}(-\alpha)+\alpha\phi_{2n-1}(-\alpha)]\Delta^{++}_{n-1}
\biggr\}\\
  \Prob( G^\symmO(N;\alpha) \le 2n+1) &=
\frac12\biggl\{ [\phi^*_{2n}(-\alpha)+\alpha\phi_{2n}(-\alpha)]
\Delta^{+-}_n
+[\phi^*_{2n}(-\alpha)-\alpha\phi_{2n}(-\alpha)]\Delta^{-+}_n
\biggr\}\\
\end{align}
where
\begin{align}
\Delta_n^{--} &= \prod_{j\ge n} N_{2j+2}^{-1} (1+\pi_{2j+2}(0))\\
\Delta_n^{++} &= \prod_{j\ge n} N_{2j+2}^{-1} (1-\pi_{2j+2}(0))\\
\Delta_n^{+-} &= \prod_{j\ge n} N_{2j+1}^{-1} (1-\pi_{2j+1}(0))\\
\Delta_n^{-+} &= \prod_{j\ge n} N_{2j+1}^{-1} (1+\pi_{2j+1}(0)).
\end{align}
We also have for $n\ge 1$,
\begin{align}
  \Prob( G^\symmO(N;1) \le 2n) &= \Delta^{-+}_n \\
\Prob( G^\symmO(N;1) \le 2n+1) &= \Delta^{++}_n \\
\end{align}
\end{lem}

The main results, Theorem \ref{thm-PV} will be obtained by analyzing the orthogonal
polynomials $\pi_k$ asymptotically. This asymptotic analysis will be carried out in Section
\ref{sec-rhp1}. But we first need the following section which will be used for the analysis
in Section \ref{sec-rhp1}.

\section{Painlev\'e V}\label{sec-PV}

In this Section, we prove various properties of a Riemann-Hilbert problem for Painlev\'e V
solution. These properties will be used in the next Section \ref{sec-rhp1} for the analysis
of orthogonal polynomials $\pi_k$, and also 
for the proof of Theorem \ref{thm-PV}.
This section is, however, independent of other sections.

\begin{figure}[ht]
 \centerline{\epsfig{file=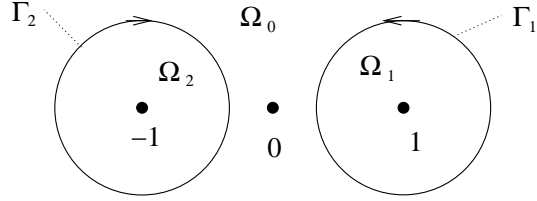, width=7cm}}
 \caption{Contours $\Gamma_1$, $\Gamma_2$ and the regions
$\Omega_1, \Omega_2, \Omega_0$}
\label{fig-Gamma}
\end{figure}

Fix $0<a<1$. Let $\Gamma_1=\{w\in\C : |w-1|=a\}$ and $\Gamma_2=\{ w : |w+1|=a\}$. We orient
the circle $\Gamma_1$ counter-clockwise, and orient the circle $\Gamma_2$ clockwise. Set
$\Gamma=\Gamma_1\cup\Gamma_2$. Let $\Omega_1, \Omega_2$ and $\Omega_0$ be the open regions
as indicated in Figure \ref{fig-Gamma}.

Let
\begin{equation}\label{e-Phi}
  \Phi(w) =\Phi(w;x,N) := \frac{(1+w)^N}{(1-w)^N} e^{-\frac12xw}.
\end{equation}
Define the $2\times 2$ matrix $V(w)=V(w;x,N)$ on $\Gamma$ by
\begin{equation}\label{e-PV1}
 \begin{cases}
V(w)=
\begin{pmatrix} 1& -\Phi(w) \\0&1
\end{pmatrix}, \quad &w\in\Gamma_1,\\
V(w)=
\begin{pmatrix} 1& 0\\ \Phi(w)^{-1}&1
\end{pmatrix}, \quad &w\in\Gamma_2.
\end{cases}
\end{equation}
Let the $2\times 2$ matrix $M(w)=M(w;x,N)=(M_{jk}(w))_{j,k=1,2}$ be the solution to the
following Riemann-Hilbert problem (RHP) :
\begin{equation}\label{e-PV2}
\begin{cases}
  M(w) \quad \text{is analytic in $w\in\C\setminus \Gamma$},\\
M_+(w)=M_-(w)V(w), \qquad w\in\Gamma,\\
M(w)= I + O(w^{-1}), \qquad \text{as $w\to\infty$}.
\end{cases}
\end{equation}
The solution $M$ shares the following properties.

\begin{prop}\label{prop-PV}
Set $\sigma_1=\bigl( \begin{smallmatrix} 0&1\\1&0
\end{smallmatrix} \bigr)$ and set
$\sigma_3=\bigl( \begin{smallmatrix} 1&0\\0&-1
\end{smallmatrix} \bigr)$.
We have the following properties :
\begin{enumerate}
\item There is a unique solution $M(w)=M(w;x,N)$ to the RHP \eqref{e-PV2}
for each $x>0$ and $N\in \N$, and the solution $M$ has the expansion
\begin{equation}\label{e-Mexpand}
  M(w)= I + \frac{M_1}{w} +\frac{M_2}{w^2} +\cdots, \qquad w\to\infty.
\end{equation}
\item $M(w)=\overline{M(\overline{w})}$, $M(w)=\sigma_1M(-w)\sigma_1$, and
$\det M(w)=1$.
\item $M(w)$ is real for $w\in\R$, and $M_1$ and $M_2$ have the form
\begin{equation}\label{e-M1definealpha}
  M_1
=\begin{pmatrix} -\alpha&\beta\\
-\beta& \alpha \end{pmatrix} , \qquad  M_2
=\begin{pmatrix} \frac12(\alpha^2-\beta^2)&\gamma \\
\gamma & \frac12(\alpha^2-\beta^2) \end{pmatrix}.
\end{equation}
Here $\alpha, \beta, \gamma$ are real constants which depend on $x$ and $N$.
\item We have
\begin{equation}\label{e-Mxderi}
  \frac{\partial}{\partial x} M = \frac14 w[M,\sigma_3]
+ \frac12 \beta\sigma_1 M.
\end{equation}
This implies, in particular,
\begin{equation}\label{e-alphabetax}
  \alpha' = \frac12 \beta^2, \qquad \beta'= \frac12(\alpha\beta-\gamma),
\end{equation}
where $'$ denotes the derivative with respect to $x$,
\item
We have
\begin{equation}\label{e-Mzderi}
  \frac{\partial}{\partial w} M = \frac{1}4 x[M,\sigma_3]
-\frac{N}{1-w^2}M\sigma_3+ \frac{N}2\frac{1}{1-w}AM
- \frac{N}2\frac{1}{1+w}\sigma_1 A\sigma_1M,
\end{equation}
where
\begin{equation}\label{e-Aformula}
  A= M(1)\sigma_3M(1)^{-1}
= \sigma_3+\frac1{N}M_1+\frac1{2N}x\beta\sigma_1(-I+M_1)
+\frac1{2N}x\gamma\sigma_1\sigma_3.
\end{equation}
\item Set $\theta(x)=\theta(x;N)=-\frac12 x\alpha(x)$.
Then $\theta$ solves the Painlev\'e V equation
\begin{equation}\label{e-thetaPV}
   (x\theta'')^2=(\theta-x\theta')(\theta-x\theta'+4(\theta')^2+4N\theta').
\end{equation}
\item Fix $0<\epsilon<1/4$. For each fixed $N\in\N$, as $x\to +\infty$,
\begin{equation}\label{e-betaasym}
  \beta(x)
= -2(-1)^NL^{(1)}_{N-1}(x)e^{-\frac12x} + O(e^{-(1-\epsilon)x}) =
\frac{-2x^{N-1}}{(N-1)!}e^{-\frac12x}(1+O(x^{-1}))
\end{equation}
and
\begin{equation}\label{e-alphaasym}
  \alpha(x) = \int_{\infty}^x 2 L_{N-1}^{(1)}(y))^2 e^{-y}dy
+ O(e^{-\frac32(1-\epsilon)x})
  = \frac{2x^{2N-2}}{((N-1)!)^2}e^{-x}(1+O(x^{-1})),
\end{equation}
where $L^{(1)}_{N-1}(x)$ is the Laguerre polynomial of degree 
$N-1$ with parameter $1$ ;
\begin{equation}
  L^{(1)}_{N-1}(x)=\sum_{j=0}^{N-1} \binom{N}{j+1}\frac{(-x)^{j}}{j!}
= \frac{1}{2\pi i} \int_{|z|=\frac12 x} \bigl( 1+\frac{x}{z}\bigr)^{N-1} 
e^{-z} \bigl( 1+\frac{z}{x}\bigr) \frac{dz}{z}.
\end{equation}
Also if we take $\Gamma_1$ and $\Gamma_2$ to be the circles
of radius $\epsilon$, centered at $1$ and $-1$, respectively,
we have for each fixed $w\in \C\setminus(\Gamma_1\cup\Gamma_2)$,
as $x\to +\infty$,
\begin{eqnarray}\label{e-Maswlarge}
  M(w) = I - \begin{pmatrix} 0& \Lambda(w,x,N) \\ \Lambda(-w,x,N) &0
\end{pmatrix}
   + O(e^{-(1-2\epsilon)x}),
\end{eqnarray}
where
\begin{equation}\label{}
  \Lambda(w,x,N) := \frac1{2 \pi i} \int_{\Gamma_1}
\Phi(s) \frac{ds}{s-w}.
\end{equation}
\item
We have
\begin{equation}\label{e-Mat0}
  M(0) = \frac12 \begin{pmatrix} \ELEN(x)^{-2}+\ELEN(x)^{2} & \ELEN(x)^{-2}-\ELEN(x)^2 \\
  \ELEN(x)^{-2}-\ELEN(x)^2 & \ELEN(x)^{-2}+\ELEN(x)^2\end{pmatrix}
\end{equation}
where $\ELEN(x)$ is defined in \eqref{e-Edefine} with $\alpha$ in \eqref{e-M1definealpha}.
\end{enumerate}
\end{prop}

\begin{proof}
\begin{itemize}
\item[(i)]
The proof of (i) is parallel to the analysis of Theorem 5.50 in \cite{DKMVZ3}. We here only
present an outline of the proof.
Let $C$ be the Cauchy operator given by
\begin{equation}
  (Cf)(w)= \frac1{2\pi i} \int_\Gamma \frac{f(s)}{s-w} ds, \qquad f\in L^2(\Gamma).
\end{equation}
Let $C_+$ be the limit from the positive side of the contour :
\begin{equation}
  (C_+f)(w)= \lim_{w'\to w} (Cf)(w'), \qquad \text{$w'$ is in the $(+)$-side of
  $\Gamma$}.
\end{equation}
Define the operator $C_V$ acting on $L^2(\Gamma)$ by
\begin{equation}\label{e-Cauchyoperator}
  (C_Vf)(w):=C_+(f(I-V^{-1})(w), \qquad f\in L^2(\Gamma).
\end{equation}
From the general theory (see e.g. Appendix I, \cite{DKMVZ2}),
if the operator $1-C_V$ is
invertible in $L^2(\Gamma)$, the function $M(w)$ defined by
\begin{equation}\label{e-solM}
  M(w)= I+ \frac1{2\pi i} \int_\Gamma
\frac{(I+(1-C_V)^{-1}(C_VI))(s)(I-V^{-1})(s)}{s-w} ds
\end{equation}
solves the RHP \eqref{e-PV2} in $L^2$ sense. But since $V$ and $V^{-1}$ are analytic on
$\Gamma$, the $L^2$ solution is indeed the classical solution of \eqref{e-PV2} (see e.g.
Proposition 5.80, \cite{DKMVZ3}). Just as in Step 1 and 2 of the proof of Theorem 5.50,
\cite{DKMVZ3}, $1-C_V$ is a Fredholm operator of index $0$, because the contour $\Gamma$ is
compact and $V$ is real analytic on $\Gamma$. Thus it is enough to show that $Ker
(1-C_V)=\{0\}$ in order to prove that $1-C_V$ is invertible. Now suppose $(1-C_V)f=0$ for
$f\in L^2(\Gamma)$. We will show that $f=0$, which will prove that $1-C_V$ is invertible.
Set
\begin{equation}\label{e-ndefine}
  n(w):= \frac1{2\pi i} \int_\Gamma \frac{f(s)(I-V^{-1})(s)}{s-w} ds.
\end{equation}
From the properties of the Cauchy operator, $n$ satisfies
\begin{itemize}
\item $n(w)$ is analytic in $\C\setminus\Gamma$, and continuous
up to the boundaries.
\item $n_+(w)=n_-(w)V(w), \qquad w\in\Gamma$.
\item $n(w)= O(\frac1{w})$ as $w\to\infty$.
\end{itemize}

Let $\Gamma_0$ be the imaginary axis, oriented from $i \infty$ to $-i\infty$. Define $N(w)$
by
\begin{equation}
\begin{cases}
  N(w)= n(w) \begin{pmatrix} 1&-\Phi(w) \\ 0&1 \end{pmatrix},
  \qquad &w\in\Omega_0, Re(w)>0, \\
  N(w)= n(w) \begin{pmatrix} 1&0\\ -\Phi(w)^{-1} & 1 \end{pmatrix},
  \qquad &w\in\Omega_0, Re(w)<0, \\
  N(w)= n(w) \qquad &w\in \Omega_1\cup\Omega_2.
\end{cases}
\end{equation}
Then $N$ satisfies
\begin{itemize}
  \item $N(w)$ is analytic in $w\in \C\setminus \Gamma_0$.
  \item $N_+(w)=N_-(w) \VN(w)$ for $w\in \Gamma_0$,
  where $\VN(w)= \begin{pmatrix} 1& -\Phi(w) \\ \Phi(w)^{-1} &0 \end{pmatrix}$.
  \item $N(w)= O(\frac1{w}e^{-\frac12x|Re(w)|})$ as $w\to\infty$.
\end{itemize}
Now consider
\begin{equation}
  a(w):=N(w)\overline{N(-\overline{w})}^T.
\end{equation}
Then $a(w)$ is holomorphic in $\C\setminus \Gamma_0$,
and hence by Cauchy's theorem and the decay
property as $w\to\infty$, we have
\begin{equation}
  \int_{\Gamma} a_+(w) dw=0.
\end{equation}
But from the jump condition of $N$,
\begin{equation}
  a_+(w)=N_+(w)\overline{N_-(-\overline{w})}^T= N_+(w)\overline{N_-(w)}^T
  = N_+(w)(\overline{\VN(w)}^{-1})^{T}\overline{N_+(w)}^T
\end{equation}
for $w\in\Gamma$.
Thus using the property $\Phi(w)^{-1}=\overline{\Phi(w)}$ for $w\in\Gamma_0$,
\begin{equation}
\begin{split}
  0 &=\int_\Gamma (a_+(w) + \overline{a_+(w)}^T) dw= \int_\Gamma N_+(w) \bigl(
  (\overline{V_0(w)}^{-1})^T+V_0(w)^{-1} \bigr) \overline{N_+(w)}^T dw \\
  &= \int_\Gamma  N_+(w) \begin{pmatrix} 0&0\\0&2\end{pmatrix}
\overline{N_+(w)}^T dw.
\end{split}
\end{equation}
This implies that $(N_{12})_+(w)=(N_{22})_+(w)=0$ for almost every $w\in\Gamma$, and hence
by Cauchy's theorem, we have
\begin{equation}
  N_{12}(w)=N_{22}(w)=0, \qquad Re(w)>0.
\end{equation}
This in turn implies, from the jump condition, that
\begin{equation}
  N_{11}(w)=N_{21}(w)=0, \qquad Re(w)<0.
\end{equation}
On the other hand, from the jump condition again,
\begin{equation}
  (N_{11})_+(w)=(N_{12})_-(w)e^{\frac12xw}\biggl(\frac{1-w}{1+w}\biggr)^N, \qquad w\in i\R.
\end{equation}
Set
\begin{equation}
  h(w):= \begin{cases} (1+w)^NN_{11}(w), \quad & Re(w)>0,\\
  (1-w)^Ne^{\frac12xw}N_{12}(w), \quad &Re(w)<0. \end{cases}
\end{equation}
Then $h$ is entire, and $h(w)w^{-N-1} \to 0$ as $w\to\infty$ since $N(w)$ is bounded for
$w\in\C$. Thus by the Liouville's theorem, $h(w)$ is a polynomial of degree at most $N$.
But then for $Re(w)<0$, $N_{12}(w)=h(w)(1-w)^Ne^{-\frac12xw}$ blows up as $w\to\infty$
unless $h=0$ identically. Therefore we obtain
\begin{equation}
\begin{split}
  &N_{12}(w)=0, \qquad Re(w)<0, \\
  &N_{11}(w)=0, \qquad Re(w)>0.
\end{split}
\end{equation}
By a similar argument, we obtain
\begin{equation}
\begin{split}
  &N_{21}(w)=0, \qquad Re(w)>0, \\
  &N_{22}(w)=0, \qquad Re(w)<0.
\end{split}
\end{equation}
Therefore we have $N(w)=0$ and hence $n(w)=0$ for all $w\in\C$.
Since by \eqref{e-Cauchyoperator} and \eqref{e-ndefine},
$n_+=C_+(f(I-V^{-1}))=C_V(f)=f$,
we obtain that $Ker(1-C_V)$ has dimension 0.
The uniqueness of the solution is standard.
The expansion \eqref{e-Mexpand} follows from \eqref{e-ndefine}.

\item[(ii)] The jump matrix $V$ have the properties,
$V(w)=\overline{V(\overline{w})}$, $V(w)=\sigma_1V(-w)^{-1}\sigma_1$
and $\det V(w)=1$ for $w\in\Gamma$.
These properties, together with the uniqueness of the solution to the RHP,
imply the results.
\item[(iii)] The realness of $M(w)$ for $w\in\R$ follows
from $M(w)=\overline{M(\overline{w})}$, which also implies
that $M_1$ and $M_2$ are real.
By expanding the both sides of $M(w)=\sigma_1M(-w)\sigma_1$ as $w\to\infty$,
we have $M_1=-\sigma_1M_1\sigma_1$ and $M_2=\sigma_1M_2\sigma_1$, and hence
\begin{equation}
  M_1
=\begin{pmatrix} -\alpha&\beta\\
-\beta& \alpha \end{pmatrix} ,
\qquad  M_2
=\begin{pmatrix} (M_2)_{11}&\gamma \\
\gamma & (M_2)_{11} \end{pmatrix},
\end{equation}
for some $\alpha, \beta, \gamma, (M_2)_{11}$. Also since $0=\log \det M = tr \log M = tr
\log (I+\frac{M_1}{w}+\frac{M_2}{w^2}+\cdots)$, as $w\to\infty$, we have $tr(M_1)=0$ and
$tr(M_2-\frac12 M_1^2)=0$. The second identity implies that
$(M_2)_{11}=\frac12(\alpha^2-\beta^2)$ and we obtain the results.
\item[(iv)]
Set $f:=Me^{-\frac14 xw\sigma_3}$. Then
\begin{equation}
\begin{cases}
f \quad \text{is analytic in $\C\setminus\Gamma$,}\\
f_+=f_-v_f(w)
\qquad \text{for $w\in\Gamma$,}\\
fe^{\frac14 xw\sigma_3}\to I \qquad \text{ as $w\to\infty$.}
\end{cases}
\end{equation}
where
\begin{equation}
\begin{cases}
v_f(w)= \begin{pmatrix} 1&-(\frac{1+w}{1-w})^N \\ 0&1 \end{pmatrix}
\qquad &w\in\Gamma_1,\\
v_f(w)= \begin{pmatrix} 1&0 \\ (\frac{1-w}{1+w})^N & 1 \end{pmatrix} \qquad
&w\in\Gamma_2.
\end{cases}
\end{equation}
Note that the jump matrix does not depend on $x$. Thus
$f'$ and $f$ satisfy the same jump condition, and hence $f'f^{-1}$ is
a entire function, where $f'=\frac{\partial}{\partial x}f$.
On the other hand, from the condition
$fe^{\frac14 xw\sigma_3}\to I$ as $w\to\infty$, we obtain
$f'f^{-1}+\frac14wf\sigma_3f^{-1}\to 0$ as $w\to\infty$.
But since $M=I+\frac{M_1}{w}+O(w^{-2})$ as $w\to\infty$,
we have $\frac14wf\sigma_3f^{-1}=
\frac14w\sigma_3+\frac14[M_1,\sigma_3]+O(w^{-1})$. Thus by Liouville's theorem,
$f'f^{-1}=-\frac14w\sigma_3-\frac14[M_1,\sigma_3]$, which is equivalent to
\begin{equation}
  M'=\frac14w[M,\sigma_3]+\frac12\beta\sigma_1M,
\end{equation}
as desired.

By taking the limit $x\to\infty$ in \eqref{e-Mxderi} and
collecting the terms of order $O(w^{-1})$, we have
\begin{equation}
  M_1'= \frac14 [M_2, \sigma_3] + \frac12 \beta\sigma_1M_1.
\end{equation}
This implies \eqref{e-alphabetax}.
\item[(v)]
Set $h=Me^{-\frac14xw\sigma_3}\bigl(\frac{1+w}{1-w}\bigr)^{\frac12N\sigma_3}$,
where
$\bigl(\frac{1+w}{1-w}\bigr)^{\frac12}$ is defined to be analytic
in $\C\setminus[-1,1]$
with the condition that it becomes $1$ as $w=iy\to 0$ satisfying $y>0$. Then
\begin{equation}
\begin{cases}
  h \quad \text{is analytic in $\C\setminus([-1,1]\cup\Gamma)$,}\\
  h_+=h_- v_h, \qquad w\in\Gamma,\\
  h_+=(-1)^Nh_-, \qquad w\in (-1,1),\\
  h e^{\frac14xw\sigma_3}\bigl(\frac{1+w}{1-w}\bigr)^{-\frac12N\sigma_3}
\to I \qquad \text{as $w\to\infty$},
\end{cases}
\end{equation}
where the interval $[-1,1]$ is oriented from the right to the left,
and the jump matrix
$v_h$ is given by
\begin{equation}
\begin{cases}
  v_h= \biggl(\begin{smallmatrix} 1&-1\\ 0&1 \end{smallmatrix} \biggr)
\qquad &w\in\Gamma_1, \\
  v_h= \biggl(\begin{smallmatrix} 1&0\\ 1&1 \end{smallmatrix} \biggr)
\qquad &w\in\Gamma_2. \\
\end{cases}
\end{equation}
Then
\begin{equation}\label{e-hdot}
  \dot{h}h^{-1} = \dot{M}M^{-1} -\frac14xM\sigma_3M^{-1}
+ \frac{N}{1-w^2}M\sigma_3M^{-1}
\end{equation}
is analytic in $\C\setminus\{-1,1\}$ with simple poles at $-1$ and $1$, where $\dot{h}$
denotes the derivative with respect to $w$. Thus, since $\dot{h}h^{-1}\to
-\frac14x\sigma_3$ as $w\to\infty$, we find that
\begin{equation}\label{e-hdot2}
  \dot{h}h^{-1} = -\frac14x\sigma_3+ \frac{A_0}{1-w}+\frac{B_0}{1+w}
\end{equation}
with some constant matrices $A_0$ and $B_0$. By taking the limits of \eqref{e-hdot} as
$w\to 1, -1$, we obtain $A_0=\frac{N}2M(1)\sigma_3M(1)^{-1}$ and
$B_0=\frac{N}2M(-1)\sigma_3M(-1)^{-1}$. Since $M(-1)=\sigma_1M(1)\sigma_1$ from (ii), we
have $B_0=-\sigma_1A_0\sigma_1$.

By combining \eqref{e-hdot} and \eqref{e-hdot2}, we obtain \eqref{e-Mzderi} with constant
matrix $A=M(1)\sigma_3M(1)^{-1}$. Now we take the limit $w\to\infty$ to \eqref{e-Mzderi}.
By collecting the terms of order $O(w^{-1})$ and noting that
$[M_1,\sigma_3]=-2\beta\sigma_1$, we have
\begin{equation}
  A+\sigma_1A\sigma_1= -\frac1{N}x\beta\sigma_1,
\end{equation}
and by collecting the terms of order $O(w^{-2})$ and noting
that $[M_2,\sigma_3]=2\gamma\sigma_1\sigma_3$, we obtain
\begin{equation}
  A-\sigma_1A\sigma_1= 2\sigma_3 + \frac2{N}M_1 + \frac1{N}x\beta\sigma_1M_1
+\frac1{N}x\gamma\sigma_1\sigma_3.
\end{equation}
These yield the second formula \eqref{e-Aformula} for $A$.

\item[(vi)]
From the second formula \eqref{e-Aformula} of $A$, we have
\begin{equation}\label{e-Aexp}
 A = \begin{pmatrix}
1-\frac1{N}\alpha -\frac1{2N}x\beta^2 &
\frac1{N}-\frac1{2N}x\beta(1-\alpha)-\frac1{2N}x\gamma \\
-\frac1{N}-\frac1{2N}x\beta(1+\alpha)+\frac1{2N}x\gamma &
-1+\frac1{N}\alpha +\frac1{2N}x\beta^2
\end{pmatrix}.
\end{equation}
On the other hand,
the first formula, $A=M(1)\sigma_3 M(1)^{-1}$, of \eqref{e-Aformula}
implies that $\det A=-1$.
Thus \eqref{e-Aexp} yield the identity
\begin{equation}
  -(N-\alpha-\frac12x\beta^2)^2- \frac14x^2\beta^2+(\beta+\frac12x(\alpha\beta-\gamma))^2=-N^2.
\end{equation}
By removing $\gamma$ and $\beta$ using \eqref{e-alphabetax}, we obtain
\begin{equation}
  -(N-\alpha-x\alpha')^2-\frac12x^2\alpha'
+\frac1{2\alpha'}(2\alpha'+x\alpha'')^2=-N^2.
\end{equation}
This becomes \eqref{e-thetaPV} if we set $\theta=-\frac12x\alpha$.

\item[(vii)]
We take $\Gamma$ as the union of circle of radius $\epsilon$,
centered at $1$ and $-1$. (We have freedom to pick a contour.)
Then from the formula of $V$, it is direct to check that 
\begin{equation}\label{e-V-1est}
  \|I-V^{-1}\|_{L^\infty(\Gamma)}\le
\frac{(2+\epsilon)^N}{\epsilon^N}e^{-(1-\epsilon)x/2}.
\end{equation}
Also by \eqref{e-Cauchyoperator},
\begin{equation}\label{e-CVest}
  \|C_V\|_{L^2(\Gamma)\to L^2(\Gamma)}\le \|C_+\|_{L^2(\Gamma)\to L^2(\Gamma)}
\|I-V^{-1}\|_{L^\infty(\Sigma^M)}
\le c_1 \|I-V^{-1}\|_{L^\infty(\Gamma)}
\end{equation}
for some constant $c_1>0$. 
Here $c_1$ can be taken to be independent of
$0<\epsilon<1/4$ from a simple scaling argument.
Hence for large enough $x$,
$\|(1-C_V)^{-1}\|_{L^2(\Gamma)\to L^2(\Gamma)}\le 1/2$,
and from \eqref{e-solM},
\begin{equation}\label{e-Mest}
\begin{split}
  M(w) &= I + \frac1{2\pi i}\int_{\Gamma} \frac{(I-V^{-1})(s)}{s-w}ds
+O\biggl(\frac{\|I-V^{-1}\|_{L^\infty(\Gamma)}^2|\Gamma|} {dist(w,\Gamma)}\biggr),
\end{split}
\end{equation}
where $|\Gamma|$ is the size of $\Gamma$. 
Also, we have 
\begin{equation}\label{e-M1asymp}
\begin{split}
  M_1 &= -\frac1{2\pi i}\int_\Gamma (I+(1-C_V)^{-1}(C_VI))(s)(I-V^{-1})(s)ds \\
&= - \frac1{2\pi i}\int_\Gamma (I-V^{-1})(s)ds
- \frac1{2\pi i}\int_\Gamma (C_VI)(s)(I-V^{-1})(s)ds \\
&\qquad - \frac1{2\pi i}\int_\Gamma
((1-C_V)^{-1}(C_V(C_VI)))(s)(I-V^{-1})(s)ds,
\end{split}
\end{equation}
and the second and the third integrals on the last line are of order
\begin{equation}
  O\bigl(\|I-V^{-1}\|_{L^\infty(\Gamma)}^2 \bigr),
\qquad O\bigl(\|I-V^{-1}\|_{L^\infty(\Gamma)}^3 \bigr),
\end{equation}
respectively.
Therefore, we obtain
\begin{equation}\label{}
\begin{split}
  \beta & =(M_1)_{12} =
- \frac1{2\pi i}\int_\Gamma (I-V^{-1})_{12}(s)ds
+ O\bigl(\|I-V^{-1}\|_{L^\infty(\Gamma)}^2 \bigr) \\
&= \frac1{2\pi i} \int_{|s-1|=\epsilon}
  \biggl(\frac{1+s}{1-s}\biggr)^Ne^{-\frac12xs}ds + O(e^{-(1-\epsilon)x})
  =: f_N(x) + O(e^{-(1-\epsilon)x}).
\end{split}
\end{equation}
After the change of variables $\frac12x(s-1) = z$, we have 
\begin{equation}
  f_N(x) = \frac{2(-1)^Ne^{-\frac12x}}{2\pi i} 
\int_{|z|=\frac12x\epsilon} \bigl( 1+\frac{x}{z} \bigr)^{N-1} e^{-z} 
\bigl(1+\frac{z}{x}\bigr) \frac{dz}{z}.
\end{equation} 
This is precisely an integral representation of 
the Laguerre polynomial $L^{(1)}_{N-1}(x)$ (see e.g., \cite{AS} 22.10.8), and 
we find
\begin{equation}
  \beta= 2(-1)^Ne^{-\frac12x} L^{(1)}_{N-1}(x) + 
O(e^{-(1-\epsilon)x}).
\end{equation} 
Similarly, from \eqref{e-M1asymp}, we have
\begin{equation}\label{}
\begin{split}
  \alpha &= (M_1)_{22}= -\frac1{2\pi i} \int_\Gamma ((C_VI)(s)(I-V^{-1})(s))_{22} ds
  +O( \| I-V^{-1}\|_{L^\infty(\Gamma)}^3) \\
  & = \frac1{(2\pi i)^2} \int_{\Gamma_1} ds \int_{\Gamma_2}dt \frac{\Phi(t)^{-1}\Phi(s)}
  {t-s} + O(e^{-3/2(1-\epsilon)x})
  =: g_N(x) + O(e^{-3/2(1-\epsilon)x}).
\end{split}
\end{equation}
By a direct computation using \eqref{e-Phi}, we find that
\begin{equation}\label{}
  g_N'(x)= \frac1{2(2\pi i)^2} \int_{\Gamma_1} ds \int_{\Gamma_2}dt
  \biggl(\frac{1-t}{1+t}\biggr)^Ne^{\frac12xt} \biggl(\frac{1+s}{1-s}\biggr) e^{-\frac12xs}
  = \frac12 (f_N(x))^2.
\end{equation}
Since $\lim_{x\to\infty} g_N(x)=0$, we find that
\begin{equation}\label{}
  g_N(x)= \int_{\infty}^x \frac12 (f_N(y))^2 dy,
\end{equation}
which proves \eqref{e-alphaasym}.

For $w\in\C$ such that $dist(w,\Gamma)\ge \frac12\epsilon$, the result
\eqref{e-Maswlarge} follows from \eqref{e-Mest}
and \eqref{e-V-1est}, \eqref{e-CVest}.
For $w$ such that $dist(w,\Gamma)< \frac12\epsilon$, we
algebraically transform the RHP so that the contour $\Gamma$ are
now the unions of circles of radius $2\epsilon$, centered at $1,-1$,
and then apply the same estimates.
By undoing the algebraic transformation and using the Cauchy's theorem,
we obtain the result \eqref{e-Maswlarge}.

\item[(viii)]
From the properties (ii),
$M(0)$ is of the form
\begin{equation}\label{}
  M(0) = \begin{pmatrix} A(x) & B(x) \\ B(x) & A(x) \end{pmatrix}
\end{equation}
for some real function $A$ and $B$ such that $A^2-B^2=1$. From the differential equation
\eqref{e-Mxderi} when $w=0$, we find
\begin{equation}\label{}
  A'=\frac12 \beta B, \qquad B'=\frac12 \beta A.
\end{equation}
Thus we have
\begin{eqnarray}
  A(x)+B(x) &=& (A(x_0)+B(x_0)) e^{\int_{x_0}^x \frac12\beta(s)ds },\\
  A(x)-B(x) &=& (A(x_0)-B(x_0)) e^{-\int_{x_0}^x \frac12\beta(s)ds }.
\end{eqnarray}
The asymptotics \eqref{e-betaasym} and \eqref{e-Maswlarge} then yield that
\begin{eqnarray}
  A(x)+B(x) &=&  e^{\int_{\infty}^x \frac12\beta(s)ds },\\
  A(x)-B(x) &=&  e^{-\int_{\infty}^x \frac12\beta(s)ds },
\end{eqnarray}
which imply \eqref{e-Mat0}
\end{itemize}
\end{proof}

In Proposition \ref{prop-PV} (i), the solution to the RHP exists for $x>0$.
Indeed it can be shown that the solution ceases to exist
when $x=0$ for $N\in\N$.
\begin{lem}
There is no solution to RHP \eqref{e-PV2} when $x=0$ for $N\in\N$.
\end{lem}
\begin{proof}
When $x=0$, $\Psi(w)=(\frac{1+w}{1-w})^N$.
Suppose there is a solution $M$ to the RHP \eqref{e-PV2}.
Let $L(w)$ be the matrix defined by
\begin{equation}\label{e-LbyM}
L(w) = \begin{cases} M(w) \begin{pmatrix} 1 & -(\frac{1+w}{1-w})^N \\
0&1 \end{pmatrix}, \qquad & w\in\Omega_1, \\
 M(w) \begin{pmatrix} 1 & 0\\ -(\frac{1-w}{1+w})^N &1
\end{pmatrix}, \qquad & w\in\Omega_2, \\
 M(w), \qquad &w\in \Omega_0.
\end{cases}
\end{equation}
Then $L(w)$ is analytic in $\C$ except that its second column has a pole of
order $N$ at $w=1$ and its first column has a pole of order $N$ at $w=-1$.
Since $L(w)\to I$ as $w\to\infty$, we find that $L$ has the form
\begin{equation}\label{e-Lform}
 L(w) = I + \sum_{j=1}^N \frac1{(1-w)^j}
\begin{pmatrix} 0 & a_j \\ 0& b_j \end{pmatrix}
+ \sum_{k=1}^N \frac1{(1+w)^k} \begin{pmatrix} c_k&0 \\ d_k& 0 \end{pmatrix},
\end{equation}
for some constants $a_j, b_j, c_k, d_k$.
From \eqref{e-LbyM} and \eqref{e-Lform}, for $w\in\Omega_1$,
\begin{equation}
  M(w)= L(w) \begin{pmatrix} 1 & (\frac{1+w}{1-w})^N \\
0&1 \end{pmatrix},
\end{equation}
and especially the $12$-entry is
\begin{equation}\label{e-567}
  M_{12}(w)= \bigl(\frac{1+w}{1-w}\bigr)^N + \sum_{j=1}^N \frac{a_j}{(1-w)^j}
+ \sum_{k=1}^N \frac{c_k(1+w)^{N-k}}{(1-w)^N}.
\end{equation}
From \eqref{e-PV2}, $M$, hence $M_{12}$, is analytic at $w=1$.
By expanding \eqref{e-567} as a Laurant series around $w=1$,
and computing the coefficient of $(1-w)^{-1}$, we obtain the condition
\begin{equation}\label{e-568}
  -(-1)^N2N + a_1 -(-1)^Nc_1=0.
\end{equation}
Similarly,
for $w\in\Omega_2$,
\begin{equation}
  M_{11}(w) = 1+ \sum_{j=1}^N \frac{a_j(1-w)^{N-j}}{(1+w)^N}
+ \sum_{k=1}^N \frac{c_k}{(1+w)^k},
\end{equation}
and by expanding $M_{11}(w)$ as a Laurant series
around $w=-1$, and calculating the coefficient of $(1+w)^{-1}$, we obtain the
second condition
\begin{equation}\label{e-570}
  -(-1)^Na_1+c_1=0.
\end{equation}
But there is no such $a_1, c_1$ satisfying \eqref{e-568} and 
\eqref{e-570} when $N\in\N$, and this is a contradiction.
Therefore, there is no solution $M$ to the RHP \eqref{e-PV2}.
\end{proof}

\section{Asymptotic analysis of orthogonal polynomials 
and the proof of Theorem \ref{thm-PV}}\label{sec-rhp1}

Set
\begin{equation}
  t=\sqrt{q}.
\end{equation}
As mentioned earlier, Theorem \ref{thm-PV}  
is obtained by taking the limit \eqref{lem-Hlimit} of the formula 
Lemma \ref{lem-disc3}.
For that purpose, we need asymptotics of $\pi_k$ and $N_k$.
In view of
Lemma \ref{lem-Hlimit} and Lemma \ref{lem-disc3}, we set
\begin{equation}\label{e-tkL}
  t=1-\frac1{2L}, \qquad k=[xL],
\end{equation}
and take the limit $L\to\infty$ in this section. 
The results are summarized in Proposition \ref{prop-opasymp} 
and Proposition \ref{prop-abasymp} below.
The asymptotics of $\Delta_n^{\pm\pm}$ are in Proposition \ref{prop-prodasymp}, 
and the proof of Theorem \ref{thm-PV} is given at the end of this section.

Let $\Sigma$ be the unit circle $\{ |z|=1\}$ in the complex plane, oriented
counter-clockwise. Let $Y$ be the solution to the following $2\times 2$ 
Riemann-Hilbert problem (RHP) :
\begin{equation}\label{e-RHPY}
\begin{cases}
 Y(w) \text{ is analytic in $w\in\C\setminus \Sigma$, and
continuous up to the boundary,}\\
 Y_+(w)= Y_-(w) \begin{pmatrix} 1& w^{-k} (1+tw)^N(1+tw^{-1})^N \\
0&1 \end{pmatrix}, \qquad w\in\Sigma,\\
 Y(w)z^{-k\sigma_3} = I+O(w^{-1}), \qquad \text{ as $w\to\infty$.}
\end{cases}
\end{equation}
Then due to the work of Fokas, Its and Kitaev (\cite{FIK}, see also \cite{BDJ}), the
orthogonal polynomials $\pi_k$ and its norm $N_k$ of 
\eqref{e-firstop} are given by
\begin{equation}\label{e-tofind}
  \pi_k(w)= Y_{11}(w), \qquad N_k=Y_{12}(0), \qquad N_{k-1}=-Y_{21}(0)^{-1}.
\end{equation}
The goal of this section is to find the asymptotics of $Y$ with 
precise error bound as $L\to\infty$ with \eqref{e-tkL}, 
and hence to find the asymptotics of $\pi_k$ and $N_k$.
We use the steepest-descent method for RHP, which was introduced by 
Deift and Zhou \cite{DZ1}.
Throughout this section, $N$ is a fixed parameter, 
while $t$ and $k$ would vary as $L$ varies. 

Define $\mone$ by
\begin{equation}
 \mone(z)= (-1)^{\frac{k}2\sigma_3} Y(z) \begin{pmatrix}
(1+tz)^N&0\\0&(1+tz)^{-N} \end{pmatrix} (-1)^{\frac{k}2\sigma_3}
\begin{pmatrix} 0&-1\\1&0 \end{pmatrix}, \qquad |z|<1,
\end{equation}
and by
\begin{equation}
 \mone(z)= (-1)^{\frac{k}2\sigma_3} Y(z) \begin{pmatrix}
z^{-k}(1+tz^{-1})^N&0\\0&z^{k}(1+tz^{-1})^{-N} \end{pmatrix}
(-1)^{-\frac{k}2\sigma_3},
\qquad |z|>1.
\end{equation}
Then $\mone$ solves the new RHP
\begin{equation}
\begin{cases}
  \mone_+(z)= \mone_-(z) \begin{pmatrix}
1&-\varphi(z) \\ \varphi(z)^{-1} &0
\end{pmatrix}, \qquad z\in\Sigma, \\
  \mone(z)= I+O(z^{-1}), \qquad \text{as $z\to\infty$,}
\end{cases}
\end{equation}
where
\begin{equation}\label{e-varphi}
  \varphi(z) := (-z)^{k}(1+tz)^N(1+tz^{-1})^{-N}.
\end{equation}
This RHP is algebraically equivalent to the RHP \eqref{e-RHPY}.

\begin{figure}[ht]
 \centerline{\epsfig{file=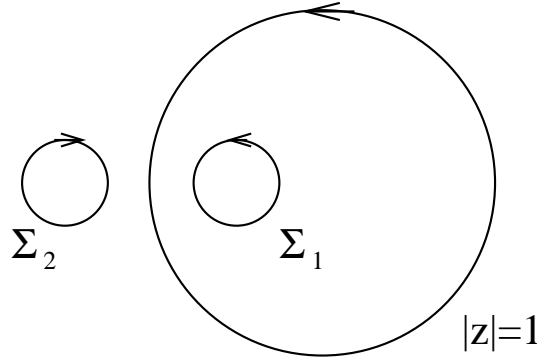, width=7cm}}
 \caption{Contours $\Sigma_1$ nd $\Sigma_2$}
\label{fig-mone}
\end{figure}

Fix $0<a<1$. Let $\Sigma_1=\{ |z+1-\frac1{2L}|=\frac{a}{2L} \}$ and $\Sigma_2=\{
|z+1+\frac1{2L}|=\frac{a}{2L} \}$. We orient the circle $\Sigma_1$ counter-clockwise, and
orient the circle $\Sigma_2$ clockwise. Note that we have plenty of freedom for the choice
of the contour. Since $0<a<1$, $\Sigma_1$, $\Sigma_2$ and $|z|=1$ have no intersection, and
the complex plane is divided into four connected regions (see Figure \ref{fig-mone}).
When $L>\frac{1+a}{2a}$, which we assume hereafter, the point $-t^{-1}$ is inside the disk
bounded by $\Sigma_2$. Define $\mtwo(z)$ by
\begin{equation}
\begin{cases}
 \mtwo(z)= \mone(z)\begin{pmatrix} 1&\varphi(z)\\0&1 \end{pmatrix},
\qquad \text{for $z$ between $\Sigma_1$ and $|z|=1$}, \\
 \mtwo(z)= \mone(z)\begin{pmatrix} 1&0\\ \varphi(z)^{-1}&1 \end{pmatrix},
\qquad \text{for $z$ in the unbounded component}, \\
 \mtwo(z)= \mone(z), \qquad \text{for $z$ inside $\Sigma_1$ and $\Sigma_2$.}
\end{cases}
\end{equation}
Since $\varphi(z)$ is analytic in $\C\setminus\{-t\}$, and $\varphi(z)^{-1}$ is
analytic in $\C\setminus\{-t^{-1}\}$, we find that $\mtwo(z)$ is analytic in
all four regions.
Moreover, by the jump condition of $\mone$ on $|z|=1$,
$\mtwo$ is analytic on $|z|=1$. Thus $\mtwo$ solves the following RHP :
\begin{equation}
\begin{cases}
  \mtwo(z) \text{ is analytic in $\C\setminus(\Sigma_1\cup\Sigma_2)$}, \\
  \mtwo_+(z) = \mtwo_-(z) \vtwo(z), \qquad z\in\Sigma_1\cup\Sigma_2, \\
  \mtwo(z) = I+O(z^{-1}), \qquad \text{as $z\to\infty$},
\end{cases}
\end{equation}
where
\begin{equation}
 \vtwo(z)= \begin{cases}
\begin{pmatrix} 1&-\varphi(z) \\ 0&1
\end{pmatrix}, \qquad z\in\Sigma_1, \\
  \begin{pmatrix} 1&0 \\ \varphi(z)^{-1}&0
\end{pmatrix}, \qquad z\in\Sigma_2.
\end{cases}
\end{equation}

Now we take the scaling
\begin{equation}
  w:= 2L(z+1).
\end{equation}
Under this map $z\mapsto w$, 
$\Sigma_1$ and $\Sigma_2$ are mapped to $\Gamma_1:= \{ w: |w-1|=a\}$ and
$\Gamma_2:= \{  w:|w+1|=a\}$, respectively. Set $\Gamma=\Gamma_1\cup\Gamma_2$. We define
\begin{equation}\label{}
  \mthree(w):= \mtwo(-1+\frac{w}{2L}).
\end{equation}
Then $\mthree$ solves the RHP
\begin{equation}
\begin{cases}
  \mthree(w)  \text{ is analytic in $w\in\C\setminus\Gamma$}, \\
  \mthree_+(w) = \mthree_-(w) \vthree(w), \qquad w\in\Gamma, \\
  \mthree(z) = I+O(w^{-1}), \qquad \text{as $w\to\infty$},
\end{cases}
\end{equation}
where
\begin{equation}\label{}
  \vthree(w)=\mthree(-1+\frac{w}{2L}).
\end{equation}

Now we need the following Lemma.

\begin{lem}\label{lem-RHPestimate}
For fixed $0<a<\frac23$, where $a$ is the radius of $\Gamma_1$ and $\Gamma_2$, we have the
following results.
\begin{itemize}
  \item[(1)] With \eqref{e-tkL}, for any $x_0>0$, there are positive constants
  $C_1, c_1, L_1$ such that
\begin{equation}\label{e-RHPestimate}
  \|\varphi(-1+\frac{w}{2L}) - \Phi(w)\|_{L^\infty(\Gamma_1)}
\le \frac{C_1}{L}e^{-c_1x},  \quad
  \|\varphi(-1+\frac{w}{2L}) - \Phi(w)\|_{L^\infty(\Gamma_2)} 
\le \frac{C_1}{L}e^{-c_1x}
\end{equation}
for all $L\ge L_1$ and $x\in[x_0,\infty)$.
  \item[(2)] For any $x_0>0$, there is a constant $C_2>0$ such that
\begin{equation}\label{}
  \| M_+\|_{L^\infty(\Gamma)}\le C_2, 
\quad \| M_+^{-1}\|_{L^\infty(\Gamma)}
  \le C_2,
\end{equation}
for all $x\in[x_0,\infty)$.
\end{itemize}

\begin{proof}
(1) By an elementary calculation, we have for any complex numbers $f,g$
\begin{equation}\label{e-easyest}
  |e^f-e^g|\le |f-g|e^{\max (Re (f),Re (g))}.
\end{equation}
For $w\in\Gamma_1$, with \eqref{e-tkL},
\begin{equation}\label{e-difference1}
  \bigl|\varphi(-1+\frac{w}{2L}) - \Phi(w) \bigr|
  = \biggl|(1-\frac{w}{2L})^{[xL]+N} -
  e^{-\frac12xw}\biggr| \biggr|\frac{1+w}{1-w}\biggl|^{N}.
\end{equation}
Setting
\begin{equation}\label{}
  f=\frac{[xL]+N}{xL}\log (1-\frac{w}{2L}),
  \qquad g= -\frac{w}{2L}
\end{equation}
for \eqref{e-easyest}, 
the difference \eqref{e-difference1} is less than or equal to
\begin{equation}\label{e-difference3}
  \bigl(\frac{2+a}{a}\bigr)^N|e^{xLf}-e^{xLg}|.
\end{equation}
Now with $k':=\frac{[xL]+N}{xL}$ and $\epsilon:= \frac{w}{2L}$,
\begin{equation}\label{e-difference2}
\begin{split}
  f-g &= k'\log(1-\epsilon)+\epsilon
  = \int_0^1 \frac{d}{ds} (k'\log(1-s\epsilon)+s\epsilon) ds \\
  &= \int_0^1 \frac{-(k'-1)\epsilon-s\epsilon^2 +(k'-1)s\epsilon\overline{\epsilon}
  +s^2\epsilon^2\overline{\epsilon}}{|1-s\epsilon|^2} ds.
\end{split}
\end{equation}
An elementary calculation yields that $|arg(w)| \le \frac{\pi}6$ for $w\in\Gamma_1$ as
$0<a<\frac23$, and hence we have $|arg(\epsilon)|\le \frac{\pi}6$ and $|arg(\epsilon^2)|\le
\frac{\pi}3$. Also, when $L\ge \frac{\sqrt{2}}{\sqrt{2}-1} \ge
\frac{a+1}{\sqrt{2}(\sqrt{2}-1)}$, we have $\frac12\le |1-s\epsilon|^2\le 2$ for
$w\in\Gamma_1$. Hence noting that $k'- 1\ge 0$, we obtain
\begin{equation}\label{}
\begin{split}
  Re(f-g) &\le \int_0^1 \bigl\{ -\frac12(k'-1) Re(\epsilon) -\frac12 s Re(\epsilon^2) +
  2(k'-1) s |\epsilon|^2 + 2s^2|\epsilon|^2 Re(\epsilon) \bigr\} ds \\
  &= -\frac12(k'-1) Re(\epsilon) - \frac14 Re(\epsilon^2) + (k'-1)|\epsilon|^2 +
  \frac23|\epsilon|^2 Re (\epsilon).
\end{split}
\end{equation}
The first term satisfies $-\frac12(k'-1) Re(\epsilon)\le 0$. For the third term, when $L\ge
\frac{16N}{x_0}$, we have $k'-1= \frac{[xL]-xL+N}{xL} \le \frac{N}{x_0L}\le \frac1{16}$,
and hence we find $(k'-1)|\epsilon|^2\le \frac18 Re(\epsilon^2)$ as $|arg(\epsilon^2)|\le
\frac{\pi}3$. It is also direct to check that the forth term has the estimate
$\frac23|\epsilon|^2 Re(\epsilon)\le \frac1{16} Re(\epsilon^2)$ if we take $L\ge
\frac{64}3$. These estimates yield
\begin{equation}\label{}
  Re (f-g)\le -\frac1{16} Re(\epsilon^2)<0
\end{equation}
when $L\ge \max(\frac{64}3, \frac{16N}{x_0})$.

Similarly, from \eqref{e-difference2}, 
using $|\epsilon|\le\frac{1+a}{L}\le
\frac{2}{L}$ and $|1-s\epsilon|^2\ge \frac12$ 
when $L\ge \frac{\sqrt{2}}{\sqrt{2}-1}$, 
we obtain
\begin{equation}\label{}
\begin{split}
  |f-g| &\le 2\int_0^1 \bigl\{ (k'-1)|\epsilon| + s|\epsilon|^2+(k'-1)|\epsilon|^2 +s^2|\epsilon|^3
  \bigr\} ds \\
  &\le (\frac{4N}{x_0}+4)\frac1{L^2} + (\frac{4N}{x_0}+\frac{16}3)\frac1{L^3}\\
  &\le 3(\frac{4N}{x_0}+4) \frac{1}{L^2}.
\end{split}
\end{equation}
Now the first inequality of \eqref{e-RHPestimate} on $\Gamma_1$ is obtained using
\eqref{e-difference3}. The second inequality of \eqref{e-RHPestimate} on $\Gamma_2$ follows
from the inequality on $\Gamma_1$ and the symmetry under $w\mapsto -\overline{w}$.

(2) From \eqref{e-solM} and \eqref{e-Cauchyoperator}, we have $M_+=I+(1-C_V)^{-1}(C_VI)$.
By Proposition \ref{prop-PV}, $(1-C_V)^{-1}$ exists for all $x>0$, and also it is easy to
check that $C_V$ and $(1-C_V)^{-1}$ are continuous in $x$. Hence from \eqref{e-solM}, $M_+$
is uniformly bounded for $x$ in a compact subset of $(0,\infty)$. Also, from the analysis
of Proposition \ref{prop-PV} (vii), $C_V$ and $(1-C_V)^{-1}$ are indeed bounded as
$x\to\infty$. Hence we obtain the uniform boundedness of $M_+$ for $x\in [x_0,\infty)$. The
boundedness of $M_+^{-1}$ follows from the boundedness of $M_+$, together with the fact
that $\det M_+=1$.
\end{proof}
\end{lem}

Set
\begin{equation}\label{e-Rdefine}
  R(w):= \mthree(w)(M(w))^{-1}.
\end{equation}
Then $R$ solves the RHP,
\begin{equation}\label{e-RHPR}
\begin{cases}
  R_+(w)=R_-(w)\vR(w), \qquad w\in\C\setminus\Sigma, \\
  R(w)= I+O(w^{-1}), \qquad \text{as $w\to\infty$},
\end{cases}
\end{equation}
where
\begin{equation}
  \vR=M_-\vthree V^{-1}(M_-)^{-1} = M_+V^{-1} \vthree M_+^{-1}.
\end{equation}
Then $v_R$ shares the following properties.

\begin{lem}\label{lem-Rasymp}
Let $C_{v_R}$ be the operator introduced in \eqref{e-Cauchyoperator} for $v_R$. For any
fixed $x_0>0$, there are positive constants $C_3, c_3, L_3$ such that
\begin{equation}\label{e-vRestimate}
  \| I-v_R^{-1}\|_{L^\infty(\Gamma)} \le \frac{C_3}{L}e^{-c_3x},
\end{equation}
uniformly in $x\in [x_0,\infty)$ for $L\ge L_3$. Also for $L$ and $x$ in the same range,
the operator $C_{v_R}$ acting on $L^2(\Gamma)$ satisfies
\begin{equation}\label{e-vRestimate2}
  \|(1-C_{v_R})^{-1}\|_{L^2(\Gamma)\to L^2(\Gamma)}\le 2,
\end{equation}
and the matrix $R$ in \eqref{e-Rdefine} is given by
\begin{equation}\label{e-Rasymp}
\begin{split}
  R(w)& =\mthree(w)(M(w))^{-1} \\
 &=
I + \frac1{2\pi i} \int_{\Gamma} \frac{I-\vR(s)^{-1}}{s-w} ds + \frac{1}{2\pi i}
\int_{\Gamma} \frac{((1-C_{\vR})^{-1}C_{\vR}I)(s)(I-\vR(s)^{-1})}{s-w} ds
\end{split}
\end{equation}
\end{lem}

\begin{proof}
Since
\begin{equation}
  I-(\vthree)^{-1}V= \begin{pmatrix} 0&\Phi(w)-\varphi(-1+\frac{w}{2L}) \\ 0&0
  \end{pmatrix},
\end{equation}
for $w\in\Gamma_1$, from Lemma \ref{lem-RHPestimate} (1), we obtain for $L\ge L_1$,
\begin{equation}
  \| I-(\vthree)^{-1}V\|_{L^\infty(\Gamma_1)} \le \frac{C_1}{L}e^{-c_1x}.
\end{equation}
A similar calculation yields the same bound for $\Gamma_2$. Therefore using Lemma
\ref{lem-RHPestimate}, we obtain for $L\ge L_1$,
\begin{equation}
  \| I-v_R^{-1}\|_{L^\infty(\Gamma)} = \| M_+ (I-(\vthree)^{-1}V) M_+^{-1}\|_{L^\infty(\Gamma)}
  \le 2C_1C_2^2\frac{1}{L}e^{-c_1x},
\end{equation}
and we obtain \eqref{e-vRestimate} (with $c_3=c_1$). For \eqref{e-vRestimate2}, note that
since the Cauchy operator $C_+$ on $L^2(\Gamma)$ is bounded, we have
\begin{equation}
   \|C_{v_R} \|_{L^2(\Gamma)\to L^2(\Gamma)} \le \frac{C'}{L} e^{-c_3x}
\end{equation}
for $L\ge L_1$ with some new constant $C'>0$. But if we take $L$ large enough so that
$\frac{C'}{L}e^{-c_3x}\le \frac{C'}{L} \le \frac12$, we have
\begin{equation}
  \| C_{v_R} \|_{L^2(\Gamma)\to L^2(\Gamma)} \le \frac12
\end{equation}
uniformly in $x\in [x_0,\infty)$. Now the result for $C_{v_R}$ follows from the Neumann
series, and \eqref{e-RHPR} follows from the general theory of RHP (cf. proof of Proposition
\ref{prop-PV} (i)).
\end{proof}

Under the sequence of transformations $Y\to \mone \to \mtwo \to \mthree$, 
the quantities
\eqref{e-tofind} of interest become
\begin{equation}\label{e-NpiR}
   N_k= \mthree_{11}(2L), \qquad N_{k-1}^{-1}=\mthree_{22}(2L),
   \qquad \pi_l(0)= -(-1)^k\mthree_{12}(2L).
\end{equation}
From \eqref{e-Rasymp}, using
\eqref{e-vRestimate} and \eqref{e-vRestimate2}, 
we find that for $x, L$ as in Lemma \ref{lem-Rasymp},
\begin{equation}\label{e-Nasymp1}
  |\mthree(2L)M(2L)^{-1} -I |\le \frac{C'}{L^2}e^{-c_3x},
\end{equation}
for some constant $C'>0$. 

Now the following estimates for $M(2L)$ follows from 
\eqref{e-solM} in the proof of Proposition \ref{prop-PV} (i)
and also the proof of Lemma \ref{lem-RHPestimate} (ii).
\begin{lem}
For any fixed $x_0>0$, let $c_3$ and $L_3$ be as in Lemma \ref{lem-Rasymp}. There are
positive numbers $C_4$ and $C_5$ such that
\begin{equation}\label{}
  \bigl| M(2L) - I -\frac{M_1}{2L} \bigr| \le \frac{C_4}{L^2} e^{-c_3x},
  \qquad  | M(2L) | \le C_5,
\end{equation}
for all $x\in [x_0, \infty)$ and $L\ge L_3$.
\end{lem}

Thus \eqref{e-Nasymp1} yields that
\begin{equation}\label{}
  \bigl| \mthree(2L) - I -\frac{M_1}{2L} \bigr| \le \frac{C_6}{L^2} e^{-c_3x},
\end{equation}
for some constant $C_6>0$. 
From this and \eqref{e-NpiR}, we obtain the asymptotics 
of $N_k$ and $\pi_k(0)$ :

\begin{prop}\label{prop-opasymp}
With \eqref{e-tkL}, for any fixed $x_0>0$, there are positive constants $C_6, c_6, L_6$
such that
  \begin{eqnarray}{}
  \bigl| N_k -1 + \frac{\alpha(x,N)}{2L} \bigr| \le \frac{C_6}{L^2} e^{-c_6x}, \\
  \bigl| N_{k-1}^{-1} -1 - \frac{\alpha(x,N)}{2L} \bigr| \le \frac{C_6}{L^2} e^{-c_6x}, \\
  \bigl| \pi_k(0) +(-1)^k \frac{\beta(x,N)}{2L} \bigr| \le \frac{C_6}{L^2} e^{-c_6x},
  \end{eqnarray}
where $\alpha$ and $\beta$ are defined in Proposition \ref{prop-PV} (iii).
\end{prop}

As in the proof Lemma 7.1 of \cite{BDJ}, it is direct to obtain 
the following result from Proposition \ref{prop-opasymp}, 
and we skip the proof.
\begin{prop}\label{prop-prodasymp}
For any fixed $x_0>0$, there are positive constants $L_7, c_7$ such that when $2n= [xL]$,
\begin{eqnarray}
  \prod_{j\ge n} N_{2j+2}^{-1} &=& \exp \biggl( \int_x^\infty \frac14 \alpha(y,N) dy \biggr)
  \bigl(1 + O(\frac1{L} e^{-c_7x})\bigr), \\
  \prod_{j\ge n} N_{2j+1}^{-1} &=& \exp \biggl( \int_x^\infty \frac14 \alpha(y,N) dy \biggr)
  \bigl(1 + O(\frac1{L} e^{-c_7x})\bigr), \\
  \prod_{j\ge n} (1\pm \pi_{2j+2}(0)) &=&  \exp \biggl( \mp \int_x^\infty \frac14 \beta(y,N)
  dy \biggr) \bigl(1 + O(\frac1{L} e^{-c_7x})\bigr), \\
  \prod_{j\ge n} (1\mp \pi_{2j+1}(0)) &=&  \exp \biggl( \mp \int_x^\infty \frac14 \beta(y,N)
  dy \biggr) \bigl(1 + O(\frac1{L} e^{-c_7x})\bigr),
\end{eqnarray}
for $x\in [x_0, \infty)$ and $L\ge L_7$.
\end{prop}

Also under the transformations $Y\to \mone\to\mtwo\to\mthree$, $\phi_k$ in
\eqref{e-phidefine} is, with the relation $z=-1+\frac{w}{2L}$, given by
\begin{equation}\label{e-phizw}
\begin{split}
   \phi_k(z)
   = \begin{cases} -(-1)^k \mthree_{12}(w), \qquad &w\in \Omega_1, \\
   z^k(1+tz)^N(1+tz^{-1})^{-N}\mthree_{11}(w), \qquad &w\in \Omega_2, \\
   -(-1)^k\mthree_{12}(w) + z^k(1+tz)^N(1+tz^{-1})^{-N} \mthree_{11}(w),
   &w\in   \Omega_0
   \end{cases}
\end{split}
\end{equation}
(recall Figure \ref{fig-Gamma}).
Hence, under \eqref{e-tkL}, from \eqref{e-Rasymp},
for any fixed $x>0$ and fixed $w\in \C$,
\begin{equation}\label{}
  \lim_{L\to\infty} \mthree(w) = M(w).
\end{equation}
(For $w$ near the contour $\Gamma$, we could take different 
radius $a$ of the contours $\Gamma_1, \Gamma_2$.) 
Therefore, \eqref{e-phizw} has the limit
\begin{equation}\label{e-phiasymp1}
  \lim_{L\to\infty} (-1)^k \phi_k(-1+\frac{w}{2L}) =
  \begin{cases}
  -M_{12}(w) \qquad &w\in \Omega_1, \\
  -M_{12}(w) + \Phi(w)M_{11}(w), \qquad &w\in \Omega_0,\\
  \Phi(w)M_{11}(w), \qquad &w\in \Omega_2,
  \end{cases}
\end{equation}
for each fixed $x>0$ and $w\in \C$, where $k=[xL]$. On the other hand, by noting that
$\phi^*_k(z)=z^k(1+\sqrt{q}z)^N(1+\sqrt{q}z^{-1})^{-N}\phi_k(z^{-1})$, we obtain from
\eqref{e-phiasymp1} that
\begin{equation}\label{e-phiasymp2}
  \lim_{L\to\infty} \phi^*_k(-1+\frac{w}{2L}) =
  \begin{cases}
  M_{11}(-w) \qquad &w\in \Omega_1, \\
  M_{11}(-w) - \Phi(w)M_{12}(-w), \qquad &w\in \Omega_0,\\
  -\Phi(w)M_{12}(-w), \qquad &w\in \Omega_2.
  \end{cases}
\end{equation}
Using the symmetry $M(w)=\sigma_1 M(-w)\sigma_1$ of Proposition \ref{prop-PV} (ii), this is
equal to
\begin{equation}\label{e-phiasymp3}
  \lim_{L\to\infty} \phi^*_k(-1+\frac{w}{2L}) =
  \begin{cases}
  M_{22}(w) \qquad &w\in \Omega_1, \\
  M_{22}(w) - \Phi(w)M_{21}(w), \qquad &w\in \Omega_0,\\
  -\Phi(w)M_{21}(w), \qquad &w\in \Omega_2.
  \end{cases}
\end{equation}

Define $M^{(1)}(w)=M^{(1)}(w;x,N)$ by
\begin{equation}\label{e-Mupper1}
  M^{(1)}(w):=
  \begin{cases}
  M(w) \qquad &w\in \Omega_1, \\
  M(w) \begin{pmatrix} 1&-\Phi(w)\\0&1 \end{pmatrix}, \qquad &w\in \Omega_0,\\
  M(w) \begin{pmatrix} 1&-\Phi(w)\\ \Phi(w)^{-1} &0 \end{pmatrix},
\qquad &w\in \Omega_2\setminus\{-1\}.
  \end{cases}
\end{equation}
In other words, from the
jump condition of the RHP \eqref{e-PV2}, $M^{(1)}$ is obtained
by taking analytic
continuation of $M(w)$ for $w\in \Omega_1$.
Then the results \eqref{e-phiasymp1} and \eqref{e-phiasymp3}
can be written in a compact form :
\begin{prop}\label{prop-abasymp}
We have with $k=[xL]$,
  \begin{eqnarray}\label{}
  \lim_{L\to\infty} (-1)^k \phi_k(-1+\frac{w}{2L}) &=& -M^{(1)}_{12}(w) \\
  \lim_{L\to\infty} \phi^*_k(-1+\frac{w}{2L}) &=& M^{(1)}_{22}(w)
\end{eqnarray}
for each $x>0$ and $w\in\C$.
\end{prop}

\subsection*{Proof of Theorem \ref{thm-PV}}

Now we prove Theorem \ref{thm-PV}.
By combining Lemma \ref{lem-Hlimit},
Lemma \ref{lem-disc3}, Proposition \ref{prop-prodasymp}
and Proposition \ref{prop-abasymp}, we obtain
for $\rho>0$ and $x>0$,
\begin{equation}
\begin{split}
   &\Prob( H^\symmO(N;\rho) \le x) \\
&\quad =
\frac12 \biggl\{ \bigl[M^{(1)}_{22}(w)-M^{(1)}_{12}(w)\bigr](\ELEN(x))^{-1}
+ \bigl[M^{(1)}_{22}(w)+M^{(1)}_{12}(w)\bigr]\ELEN(x)\biggr\}
\FLEN(x),
\end{split}
\end{equation}
where
\begin{equation}
   w= \frac{2}{\rho}-1.
\end{equation}
Thus we set
\begin{equation}
   \aLEN(x,\rho) = M^{(1)}_{22}(\frac{2}{\rho}-1; x),
\qquad \bLEN(x,\rho) = M^{(1)}_{12}(\frac{2}{\rho}-1; x).
\end{equation}
From the RHP for $M$, $\aLEN(x,\rho), \bLEN(x,\rho)$ are analytic in 
$x>0, \rho>0$, and $\ELEN(x), \FLEN(x)$ are analytic in $x>0$.

The properties (i), (ii) of Theorem \ref{thm-PV}
are given in Proposition \ref{prop-PV} (iv), (vi)
and (vii).

For the property (iii), we note that it is direct to
check that when $M$ satisfies the differential
equations (vi), (v) of Proposition \ref{prop-PV},
$M^{(1)}$ also satisfies the \emph{same} differential equations.
This implies the property (iii).

The asymptotics (iv) of $\aLEN, \bLEN$ follows from \eqref{e-Maswlarge},
and the definition of $M^{(1)}$.

The asymptotics \eqref{e-abatrho0} in (v) follows from the
normalization condition of the RHP : $M(w)\to I$ as $w\to\infty$,
and \eqref{e-abatrho2} is obtained from Proposition \ref{prop-PV} (viii)
and the definition of $M^{(1)}$.

Now we compute \eqref{e-abatrhoinfty}.
With $x=y\rho$ and $w=\frac{2}{\rho}-1$, we have
\begin{equation}\label{e-proof1}
  \Phi(w;x)= e^{-y}e^{\frac12 y\rho} \frac1{(\rho-1)^N} = O(e^{\frac12 y\rho}).
\end{equation}
As $\rho\to\infty$, using \eqref{e-aasymp},
with the change of variables $s=1-(2/\rho)u$,
\begin{equation}
\begin{split}
  \aLEN(y\rho, \rho) &= \Phi(\frac{2}{\rho}-1; y\rho) \bigl(
\Gamma_1 (-\frac{2}{\rho}+1, y\rho) + O(e^{-(1-2\epsilon) y\rho}) \bigr)\\
&=
\frac{1}{2\pi i} \int_{|u|=1/2} e^{y(u-1)}
\bigl( \frac{\rho-u}{\rho-1}\bigr)^N \frac{du}{u^N(u-1)}
+ O(e^{-(\frac12-2\epsilon)y\rho}) \\
&= \frac{1}{2\pi i} \int_{|u|=1/2} e^{y(u-1)} \frac{du}{u^N(u-1)}
\bigl( 1+ O(\rho^{-1}) \bigr) + O(e^{-(\frac12-2\epsilon)y\rho}).
\end{split}
\end{equation}
But the function
\begin{equation}
   g(y) := \frac{1}{2\pi i} \int_{|u|=1/2} e^{y(u-1)} \frac{du}{u^N(u-1)}
\end{equation}
satisfies $g'(y)= e^{-y} \frac{y^{N-1}}{(N-1)!}$ and $g(0)=0$.
Hence $g(y)= P(N,y)$, the incomplete Gamma function,
and we obtain the first of \eqref{e-abatrhoinfty}.
The second of \eqref{e-abatrhoinfty} follows from \eqref{e-basymp}
and  \eqref{e-proof1}.

\section{Limiting distributions as $N\to\infty$}\label{sec-limit}

In Theorem 4.2 of \cite{BR3}, the authors computed the limiting distributions of the last
passage percolation time $G^\symmO(N;\alpha)$ of \eqref{e-Ggeom} as $N\to\infty$ for
various values of $\alpha$. 
Similar results are also obtained in \cite{BR2}
for a Poisson percolation model 
with a symmetry condition, which also has interpretations 
as the longest increasing subsequence of random involutions.
In this section, we take similar limit for $H^\symmO(N;\rho)$.
The results are such that we take the formal limit Lemma \ref{lem-Hlimit} in Theorem 4.2 of
\cite{BR3} assuming that the two limits $N\to\infty$ and $L\to\infty$ interchange. The
functions $F_1, F_4$ and $F^\symmO$ are defined in \cite{BR3, BR2}.

\begin{thm}\label{thm-limit}
For each fixed $x\in \R$ and $\rho$,
\begin{equation}\label{}
  \lim_{N\to\infty} \Prob \biggl( \frac{H^\symmO(N;\rho)-4N}{2^{4/3}N^{1/3}} \le x\biggr) =
  \begin{cases} F_4(x), \qquad &0\le \rho <2, \\
  F_1(x), \qquad &\rho=2, \\
  0, \qquad &\rho>2.
  \end{cases}
\end{equation}
Also, for each fixed $x\in\R$ and $w\in\R$,
\begin{equation}\label{}
  \lim_{N\to\infty} \Prob \biggl( \frac{H^\symmO(N;\rho)-4N}{2^{4/3}N^{1/3}} \le x\biggr) =
  F^\symmO(x;w), \qquad \rho=2-\frac{2^{5/3}w}{N^{1/3}}
\end{equation}
\end{thm}

From the formula in Theorem \ref{thm-PV}, 
this result can be obtained by applying the Deift-Zhou 
steepest-descent method to the Riemann-Hilbert problem \eqref{e-PV2}.
The analysis is analogous to that of \cite{BDJ, BR2}, 
and we do not provide any details here. 
Steepest-descent analysis for an extension of the RHP \eqref{e-PV2}, 
which includes the analysis of the above 
Theorem \ref{thm-limit} as a special case, 
will be carried out in a later publication, \cite{BTASEP}.

\section{Correlation functions}\label{sec-corr}

In this last section, we give some remarks on the correlation functions 
for the general interpolating ensemble  
with the density \eqref{e-dens} with $\beta=1$,
\begin{equation}\label{e-corrint}
  p(\xi_1,\cdots, \xi_N;A;w)=  \frac1{Z_A} \prod_{1\le i<j\le N} (\xi_i-\xi_j)
  \prod_{j=1}^N w(\xi_j) e^{A(-1)^j\xi_j},
\end{equation}
on $\R^N_{ord}:=\{\xi_N\le \xi_{N-1}\le \cdots\le \xi_1\}$ ($N$ is even).
It has been known that 
the correlation functions 
for the orthogonal ensemble (when $A=0$) 
and the symplectic ensemble (when $A=+\infty$)
can be expressed in terms of the Pfaffian, 
or the square root of the determinant, of an antisymmetric 
matrix (see e.g., \cite{Mehta, TracyWidomcluster}).
For \eqref{e-corrint} with general $A$, Rains (\cite{rains:corr}) 
computed the correlation functions and showed 
that it is again expressible in terms of Pfaffians 
but of a different matrix.
In this section we remark that the result of Rains can be obtained 
from the argument of Tracy and Widom \cite{TracyWidomcluster} 
after a minor change.

In \cite{TracyWidomcluster}, Tracy and Widom developed a systematic 
method to express various correlation functions of orthogonal and symplectic
ensembles in terms of certain Fredholm determinants.
Especially in Section 9 of \cite{TracyWidomcluster}, Tracy and Widom 
expressed the correlation functions of general orthogonal ensemble.
The computation below for the above ensemble \eqref{e-corrint} 
is identical to that of Section 9 of \cite{TracyWidomcluster} 
except the change of the asymmetric factor $\epsilon \to \epsilon_A$.


We assume that $N$ is even in the below. 
When $N$ is odd, one needs some change of the formulas.
The starting point is the following identity in Remark 7.6.1, \cite{BR1} :
\begin{equation}\label{e-interPfaff}
  e^{A\sum_{j=1}^N(-1)^j\xi_j} 
= \Pf \bigl( \sgn(\xi_j-\xi_k)e^{A|\xi_j-\xi_k|} \bigr)_{1\le j,k\le N}.
\end{equation}
This can be checked by noting that the Pfaffian on the 
right-hand-side is the square root
of the determinant $D_N$ of the matrix 
$\bigl( \sgn(\xi_j-\xi_k)e^{A|\xi_j-\xi_k|} \bigr)_{1\le j,k\le N}$, 
and by finding the relation $D_N=e^{2A(\xi_{N-1}-\xi_N)}D_{N-2}$
using proper row and column operations.
From \eqref{e-interPfaff}, for any bounded
function $f$, we have
\begin{equation}\label{e-corr1}
\begin{split}
  &Z_A\cdot \int_{\R^N_{ord}}
p(\xi_1,\cdots,\xi_N; A;w)\prod_{j=1}^N (1+f(\xi_j)) d\xi_1\cdots d\xi_N  \\
  &\qquad =
  \int_{\R^N_{ord}} \Pf \bigl( \sgn(\xi_j-\xi_k)e^{A|\xi_j-\xi_k|} \bigr)_{1\le
  j,k\le N} \det\bigl( \xi_k^jw(\xi_k)(1+f(\xi_k)) \bigr)_{1\le j,k\le N} d\xi_1\cdots d\xi_N
\end{split}
\end{equation}

We also need the following result of de Bruijn
(equations (4.6)-(4.8) of \cite{deBruijn}) :
given a measure space $(X, \mu)$,
for an anti-symmetric function $s(x,y)=-s(y,x)$ on $X\times X$,
\begin{equation}\label{e-deBruijn}
\begin{split}
 &\frac1{N!} \int_{X^N} \Pf ( s(x_i,x_j))_{1\le i,j\le N}
\det ( \phi_i(x_j))_{1\le i,j\le N} d\mu(x_1)\cdots d\mu(x_N) \\
 &\quad = \Pf \biggl( \int_X\int_X \phi_i(x)s(x,y)\phi_j(y)
d\mu(x) d_\mu(y) \biggr)_{1\le i,j\le N}
\end{split}
\end{equation}
for a sequence of functions $\phi_j$ on $X$.
This is a generalization of (1.4) of \cite{TracyWidomcluster}
where the authors took the special case when $s(x,y)=\sgn(x-y)$.
Now with $s(x,y)=\sgn(x-y)e^{A|x-y|}$ and $\phi_j(x)=x^j w(x)(1+f(x))$,
the square of \eqref{e-corr1} is equal to
\begin{equation}\label{e-corr2}
   \det \biggl( \int_{\R} \int_{\R} \sgn(x-y) e^{A|x-y|} x^jy^k w(x)w(y)
(1+f(x))(1+f(y)) dx dy \biggr)_{0\le j,k \le N-1}.
\end{equation}
Here the $N!$ term in \eqref{e-deBruijn} disappears since $\R^N_{ord}$
is the ordered set $\{ \xi_N\le\cdots \le \xi_1\}$
and we take $X=\R$.
This formula is same as the second displayed formula of section 9
of \cite{TracyWidomcluster} with the modification that
$\epsilon(x-y)$ (which is $\sgn(x-y)$)
is replaced with $\sgn(x-y) e^{A|x-y|}$.

The rest of argument is same as section 9, \cite{TracyWidomcluster}
except that
the operator $\epsilon$ whose kernel is $\sgn(x-y)$ in \cite{TracyWidomcluster}
is now changed to the operator $\epsilon_A$, defined by
\begin{equation}
  (\epsilon_A h)(x) = \int_{\R} \epsilon_A(x-y) h(y) dy,
\qquad \epsilon_A(x-y):= \sgn(x-y) e^{A|x-y|},
\end{equation}
for a suitable class of functions $h$.
With this modification, the result analogous to
(3.3), (9.1) of \cite{TracyWidomcluster} is the following.

\begin{lem}(Theorem 1.1 of \cite{rains:corr})
Set $\psi_j(x)=p_j(x)w(x)$ where $p_j(x)$, $j=0,1,\cdots$, is an arbitrary
sequence of polynomials of exact degree $j$.
Let $M$ be the matrix
\begin{equation}
  M = \biggl( \int_{\R} \int_{\R} \epsilon_A(x-y) \psi_j(x) \psi_k(y) dx dy
\biggr)_{0\le j,k\le N-1}
\end{equation}
and set $M^{-1}=(\mu_{jk})$.
Then we have
\begin{equation}
 \int_{\R^N_{ord}}
p(\xi_1,\cdots,\xi_N; A)\prod_{j=1}^N (1+f(\xi_j)) d\xi_1\cdots d\xi_N
= \sqrt{\det ( 1+ K_N^{(A)} f)}
\end{equation}
where the operator $K_N^{(A)}$ has the $2\times 2$ matrix kernel
\begin{equation}\label{e-Kkernel}
  K_N(x,y;A) = \begin{pmatrix}
S_N(x,y;A) & S_ND(x,y;A) \\
IS_N(x,y;A) - \epsilon_A(x-y) & S_N(x,y;A)
\end{pmatrix}
\end{equation}
and
\begin{eqnarray}\label{e-Skernel}
  S_N(x,y;A) &=&
-\sum_{j,k=0}^{N-1} \psi_j(x)\mu_{jk} (\epsilon_A \psi_k)(y), \\
\label{e-ISkernel}
 IS_N(x,y;A) &=&
-\sum_{j,k=0}^{N-1} (\epsilon_A \psi_j)(x)\mu_{jk} (\epsilon_A \psi_k)(y),\\
\label{e-SDkernel}
S_ND(x,y;A) &=&
\sum_{j,k=0}^{N-1} \psi_j(x)\mu_{jk} \psi_k(y).
\end{eqnarray}
\end{lem}

The matrix elements in the above determinant 
for general $A$ is significantly simplified 
for the Laguerre case, $w(x)=e^{-x}1_{x\ge 0}$ by \cite{ForresterR2}, 
which is an extension for general $A$ of the results 
for Laguerre orthogonal and symplectic ensembles 
(see e.g. \cite{AFNvM, Wskew}).

When $A=0$, $\epsilon_A=\epsilon$ in the notation
of \cite{TracyWidomcluster}, and the kernel \eqref{e-Kkernel} is equal to
(9.1) of \cite{TracyWidomcluster}, which is the $\beta=1$ orthogonal ensemble.
On the other hand, when $A\to +\infty$, we do not have a proof that
$K_N(x,y;A)$ converge to the kernel (8.1) of \cite{TracyWidomcluster}
for the $\beta=4$ symplectic ensemble.
However, we note that for smooth $h$ which decays fast at $\pm\infty$,
integrations by parts yield that
\begin{equation}
\begin{split}
   (\epsilon_Ah)(x) &= -\frac2{A^2} h'(x) + \frac1{A^2} (\epsilon_A h'')(x) \\
&= -\frac2{A^2} h'(x) - \frac2{A^4} h^{(3)}(x)
+ \frac1{A^4} (\epsilon_A h^{(4)})(x)
= \cdots.
\end{split}
\end{equation}
Thus when $A\to +\infty$, it seems that the main contribution to
$(\epsilon_A h)$ comes from $-h'$.
If we replace $(\epsilon_A \psi_k)$ in \eqref{e-Skernel}-\eqref{e-SDkernel}
above by $-\psi_k'$ and drop the term $\epsilon_A(x-y)$,
\eqref{e-Kkernel} is equal to (8.1) of
\cite{TracyWidomcluster} if the notations $IS_N$ and $S_ND$ there are
exchanged.
This seems to be an indication that the kernel $K_N(x,y;A)$ actually converges
to the kernel (8.1) of \cite{TracyWidomcluster},
the $\beta=4$ symplectic ensemble, as $A\to +\infty$.
Nevertheless, Proposition \ref{prop-inter} shows that the determinant 
$\det (1+K^{(A)}_Nf)$ converges to the corresponding determinant 
for $\beta=4$ symplectic ensemble as $A\to\infty$ for a proper class of 
functions $f$.


We finish this section with some properties of the operator $\epsilon_A$
which can be checked easily.
Let $h$ be a function in the Schawrz class.
\begin{itemize}
\item
Let $g=\epsilon_Ah$. Then $g''-A^2g=2h'$.
\item
If $h$, in addition to the smooth and decay conditions,
satisfies $\int_{\R} h(x)ds=0$, we have
\begin{equation}
  (\epsilon_A^{-1}h)(x)
= \frac12 h'(x) - \frac{A^2}{2} \int_{-\infty}^x h(t)dt.
\end{equation}
\end{itemize}

\bibliographystyle{plain}
\bibliography{paper11}


\end{document}